\pdfoutput=1
\documentclass[oneside,phd,11pt]{ucl_thesis}

\usepackage[T1]{fontenc}

\usepackage{nomencl}
\makenomenclature

\usepackage{url}
\usepackage{hyperref}
\hypersetup{pdfborder={0 0 0}}
\usepackage[printonlyused]{acronym}
\usepackage{textcase}
\usepackage{graphicx}
\usepackage{epsfig}
\usepackage{amssymb}
\usepackage{amsmath}
\usepackage{amsfonts}

\usepackage{caption}
\usepackage{subcaption}
\usepackage{verbatim}

\usepackage{color}
\usepackage{array}
\usepackage{multirow}

\usepackage{graphicx}
\usepackage{caption}
\usepackage{amsthm}
\usepackage{acronym}
\usepackage[section]{algorithm}
\usepackage{algorithmic}
\usepackage{color, colortbl}
\definecolor{anti-flashwhite}{rgb}{0.95, 0.95, 0.96}
\definecolor{ashgrey}{rgb}{0.7, 0.75, 0.71}
\definecolor{cadetgrey}{rgb}{0.57, 0.64, 0.69}
\definecolor{gainsboro}{rgb}{0.86, 0.86, 0.86}
\definecolor{isabelline}{rgb}{0.96, 0.94, 0.93}

\usepackage{subcaption}
\usepackage{multirow}

\usepackage{enumitem}

\newcolumntype{$}{>{\global\let\currentrowstyle\relax}}
\newcolumntype{^}{>{\currentrowstyle}}

\makeatletter
\renewcommand{\@chapapp}{Section}
\makeatother

\title{Engineering data-driven solutions for future mobility: perspectives and challenges}

\author{
\textit{Contributors:} Daphne Tuncer\thanks{Laboratoire Ville Transport Mobilite, Ecole des Ponts ParisTech, France}, Oytun Babacan\thanks{Grantham Institute for Climate Change, Imperial College London, UK}, Raoul Guiazon\thanks{Department of Physics and Astronomy, University of Leeds, UK}, Halima Abu Ali\thanks{Work carried out as part of the Data Science and Digitalisation of the Energy Sector lecture, SEF MSc programme, Imperial College London, UK}, Josephine Conway\footnotemark[4], Sebastian Kern\footnotemark[4], Ana Teresa Moreno\footnotemark[4], Max Peel\footnotemark[4], Arthur Pereira\footnotemark[4], Nadia Assad\footnotemark[4], Giulia Franceschini\footnotemark[4], Margrethe Gjerull\footnotemark[4], Anna Hardisty\footnotemark[4], Imran Marwa\footnotemark[4], Blanca Alvarez Lopez\footnotemark[4], Ariella Shalev\footnotemark[4], Christopher D'Cruz Tambua\footnotemark[4], Hapsari Damayanti\footnotemark[4], Paul Frapart\footnotemark[4], Sacha Lepoutre\footnotemark[4], Peer Novak\thanks{ School of Environmental Sciences, University of East Anglia, UK}
}

\date{\textit{March 2022}} 
\begin{document}
\maketitle

\newpage

\newpage

\chapter*{Abstract}
\label{ref:abstract}

The automotive industry is currently undergoing major changes. These include a general shift towards decarbonised mode of transportation, the implementation of mobility as an end-to-end service, and the transition to vehicles that increasingly rely on software and digital tools to function. Digitalisation is expected to play a key role in shaping the future of mobility ecosystems by fostering the integration of traditionally independent system domains in the energy, transportation and information sectors. This report discusses opportunities and challenges for engineering data-driven solutions that support the requirements of future digitalised mobility systems based on three use cases for electric vehicle public charging infrastructures, services and security. 

\cleardoublepage \tableofcontents

\chapter{Introduction}
\label{sect:intro}

The automotive industry is going through a paradigm shift that is shaping the future of mobility. There are several key transformations happening in the automotive domain, which includes transitioning away from dependency on oil products and introducing innovative technologies for commuting. These changes are accelerated in the last few years by new political agendas at the local and international levels that promote climate change mitigation and clean-air policy goals.

Main changes in automotive are manifesting in three emerging trends: 1) sustainability, 2) mobility-as-a-service, and 3) connected and autonomous vehicles. These changes range the full spectrum of mobility, from changes operating at the vehicle level, specifically under two vectors, \textit{i.e.,} 1) the push for low emission technologies, including vehicle fleet electrification, and 2) the importance of \textit{softwarisation} as part of a vehicle's functionality (\textit{e.g.,} through embedded sensors, on-board software, \textit{etc.}), to end user delivered digital services. 

\begin{figure}[hbt!]
\centering
\includegraphics[trim = 1cm 1cm 1cm 1cm, clip, scale=0.375]{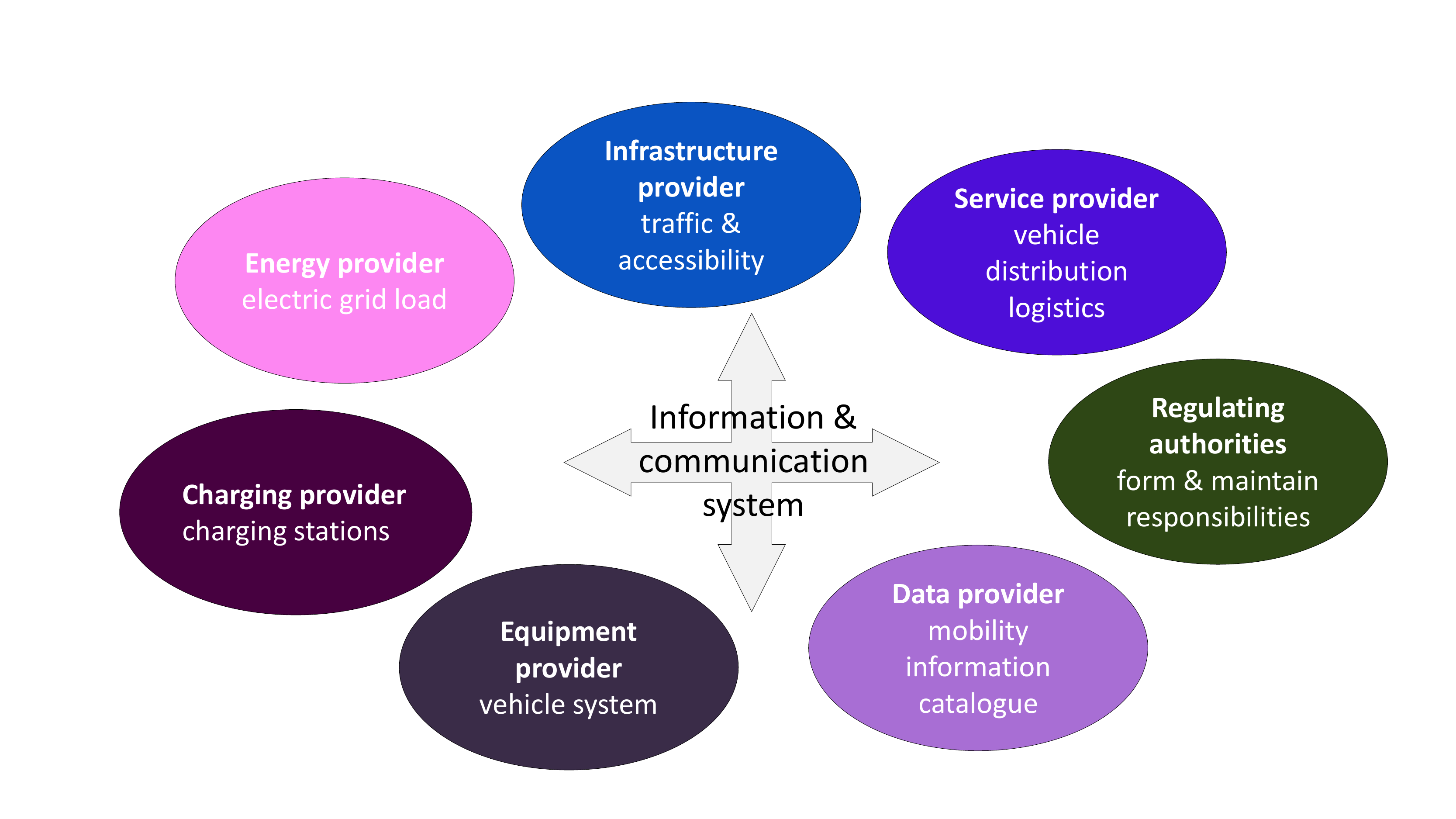}  
\vspace{-2mm}
\caption{Integration of stakeholders through ICT.}
\label{fig:stakeholdersICT}
\end{figure}

A key aspect of the implementation of these transformations is the convergence between traditionally independent system domains in the energy, transportation and digital sectors, where Information and Communication Systems (ICSs) act as a focal point. Figure \ref{fig:stakeholdersICT} depicts key domains where stakeholders participate in the delivery of electric mobility services through an integrated digitalised ecosystem, including (hardware/software) resource vendors, infrastructure operators and service providers. 

\textit{Data} plays a crucial role in the realisation of inter-operable system domains by acting as the medium interfacing system operations. In fact, today the mobility ecosystem is witnessing an unprecedented volume of data \cite{obstfeldReport}, which brings novel opportunities but also challenges to the transportation sector. Domains and their associated stakeholders can harness enriched information from the available data that can be used to optimise infrastructure planning, investment and operation, as well as, to contribute to improving user experience in terms of convenience, accessibility and added value services. At the same time, critical issues associated with the management of data-driven approaches need to be positioned as central considerations to the development of future mobility solutions, specifically issues around data security and privacy. 

In this report we discuss opportunities and challenges for engineering data-driven solutions that have the potential to support future connected and automated mobility. The report does not aim at providing definite answers to the issues that the development of data-driven approaches will inevitably bring in, but instead focuses on exposing perspectives and insights based on three use cases that we have been working on in the last few years, namely the planning of a public charging infrastructure for electric mobility (Section \ref{sect:infrastructure}); the development of added-value electric vehicle charging services (Section \ref{sect:services}); and security and data protection for connected and automated mobility (Section \ref{sect:securityPrivacy}).

\chapter{E-mobility demand: charging infrastructure planning}
\label{sect:infrastructure}

A wide-spread delivery of electric mobility (e-mobility) services requires cost effective deployment of a charging infrastructure that consists large connected networks of charging stations. The electric vehicle drivers today can access charging stations in three different ways: \textit{i)} private charging stations that are located at home or in private business premises and whose access is limited to an exclusive set of drivers; \textit{ii)} publicly accessible stations usually located at special point of interests such at commercial parks, shopping centers, train stations, \textit{etc.} whose access is often restricted to customers and business operating hours; and \textit{iii)} public charging stations that are located in public parking areas or roadside parking space with very few restrictions for public access.  

Understanding the needs and habits of e-mobility users in terms of travelling and charging helps informing the deployment plan of these charging networks. Tools based on behavioural modelling, \textit{e.g.,} \cite{ferrara2019multimodal} and \cite{yang2018user}, and predictive analytics, \textit{e.g.,} \cite{straka2020predicting} and \cite{birkel2020challenges}, for instance, assist practitioners in assessing user accessibility challenges (\textit{e.g.,} where to deploy new stations), understanding economic challenges (\textit{e.g.,} deployment and operational costs) and sustainability concerns (\textit{e.g.,} source of energy supply). They can also be employed to test mechanisms and create prototypes that would encourage better e-mobility practices, for instance through reward schemes such as the one presented by Pluntke in \cite{pluntkeDissertation14}.

In this context we have been investigating data-based opportunities for expansion planning of e-mobility charging infrastructure by taking the Greater London area in the United Kingdom as a use case. In this section, we report on three lines of work we engaged in that domain. We developed a data model that builds on top of the design of a demand-to-supply system to characterise the key features of e-mobility (Section \ref{sect:demandSupplySystem}). In addition we ran data collection studies in order to capture the specifics of e-mobility users' habits in the Greater London area (Section \ref{sect:demandCaseStudy}). Finally, we developed a framework to estimate the demand factors that affect how public charging infrastructure would be accessed by e-mobility used (Section \ref{sect:demandModel}).  

\section{Demand versus supply - a systems view}
\label{sect:demandSupplySystem}

\begin{figure}[hbt!]
\centering
\includegraphics[trim = 0cm 2cm 0cm 0cm, clip, scale=0.40]{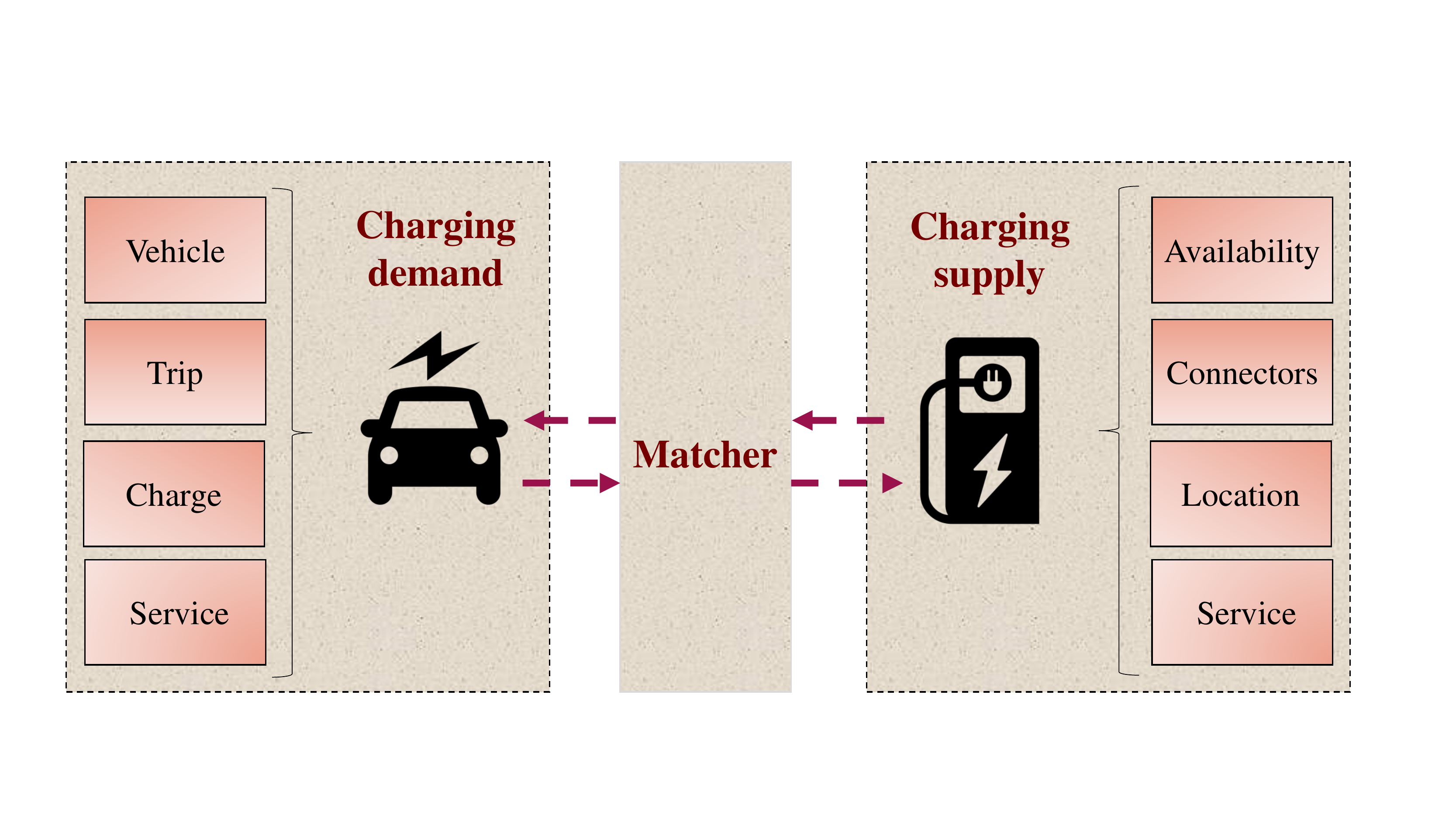}  
\caption{A demand \textit{vs.} supply system for e-mobility infrastructure usage.}
\label{fig:system}
\end{figure}

The usage on an e-mobility infrastructure can functionally be represented as a \textit{demand-to-supply matching} system where e-mobility users form a charging \textit{demand} that is satisfied by the operations of a \textit{supply} of electric charging stations. The system can be used to evaluate the impact of the demand on the supply (and \textit{vice versa}) by modelling features of travelling and charging habits on one hand, and of the network of charging stations on the other. Fig. \ref{fig:system} illustrates the functionality of the system that is decomposed into three main components: 1) e-mobility charging demand, 2) e-mobility charging supply, 3) demand-to-supply matching via vehicle-to-station allocation procedures, \textit{e.g.,} \cite{lam2014electric}. 

Various aspects of the demand and the supply can impact how the infrastructure is used. These aspects can be categorised in classes of features representing the different sub-components of the system, as shown in Table \ref{tab:features}. Features are described through attributes, \textit{i.e.,} a charge is described in terms of charging type, a battery level of the vehicle, a charging behaviour of the user and/or the charger, \textit{etc.}, whose value can be tuned to represent different characteristics of the demand and/or the supply. It is, for instance, possible to test the influence of different trip profiles on the need for charging stations in a geographical area by considering journey characteristics that affect energy consumption of the vehicle, such as the type of road, the route topology, the presence of traffic lights \textit{etc.}

\newcolumntype{g}{>{\columncolor{anti-flashwhite}}c}
\begin{table}[h]
    \caption{Classes of demand and supply features.}
\small
    \begin{center}
\begin{tabular}{|g p{4.3cm} g p{4.3cm}|}
  \hline
\rowcolor{isabelline}
  \multicolumn{2}{|c}{\textbf{Demand}} & \multicolumn{2}{c|}{\textbf{Supply}} \\
 \hline
  Vehicle  & Make, charging socket type(s), \textit{etc.} & Availability & Station real-time occupancy status \\
    Trip    & characteristics of the journeys in terms of distance, types of roads, topography, \textit{etc.}      &  Connector & sockets and plugs available at the station\\
 Charge   & both in terms of type of charge (rapid, fast, super-fast) and charging behaviour (battery level at which the vehicle is charged, battery level at which the charge is terminated, \textit{etc.})  & Location & geo-coordinates of the post, type of premises where the station resides (on-street, in a supermarket, at a station, \textit{etc.})\\ 
Service   & \textit{i.e.,} indication of service requirements such as subscription to a specific charging provider, requests for a specific charging time and location, \textit{etc.}  & Offered service at location & charge rate (if any), option to reserve a charge slot, restricted access, \textit{etc.}\\
\hline
\end{tabular}
\end{center}
\label{tab:features}
\end{table}

The design is based on a plug-in and -out model that makes each sub-component independent. A modular design facilitates the integration of multiple sets of data as part of a single platform, and as such enables the implementation of customised test scenarios. 

\section{Modeling e-mobility user demand: Greater London case study}
\label{sect:demandCaseStudy}

Features of the e-mobility demand (\textit{e.g.,} how often a vehicle is charged, its typical journey, \textit{etc.}) can be extracted using different data sources. In 2019, we ran a data collection study as part of the EVOLVE project \cite{EvolveProject} to qualitatively characterise the travel and charging behaviours of electric vehicle drivers in Greater London, UK.

To set the context of this case study, we provide here some statistics on the Greater London region in terms of its administrative structure, urban environment, existing car fleet and charging infrastructure. This region is organised in $32$ boroughs, \textit{i.e.,} inner and outer London boroughs, and the City of London that has a specific structure \cite{LondonCouncilsStat}. The urban environment is spread on a surface of $1572$ $km^{2}$. At the end of 2019, the number of licensed ultra low emission vehicles\footnote{Ultra low emission vehicles are vehicles that are reported to emit less than 75g of carbon dioxide (CO2) from the tailpipe for every kilometre travelled \cite{ULEVsStat}.} (ULEV) in Greater London was $38,632$ \cite{ULEVsStat}, including $16,740$ licensed ULEV in inner London boroughs and $21,827$ in outer London boroughs, accounting for 1.3\% of all licensed vehicles in the region \cite{DfTLicensingStat}. 

As of November 2020, Greater London counted for 26\% the UK charge point connectors\footnote{The UK was covered with $20,964$ charging stations according to Zap-Map \cite{zapmapStat}, the UK reference website about electric vehicles} \cite{tfl1120Report}. These stations are operated by several charging network operators \cite{zapmapNetw}.

\subsection{Survey of the preferences and habits of electric vehicle drivers}
\label{sect:survey}

We administrated a survey on travelling and charging habits among the community of electric vehicle (EV) drivers in Greater London in Spring 2019. Our objective was to investigate if it was possible to extract patterns for drivers in the city that could further be used as inputs to the sub-components of the demand model presented in Section \ref{sect:demandSupplySystem}. In total the survey consisted in $50$ questions related to four main aspects of e-mobility: \textit{i)} vehicle ownership, \textit{ii)} travel habits, \textit{iii)} charging habits, and \textit{iv)} charging service preferences. The obtained dataset includes a total of $166$ samples from which $139$ are suitable for further analysis\footnote{Samples with missing answers or completed in a time to short to be reasonable were discarded.}.

\subsubsection{Vehicle characteristics}

\begin{figure*}[!t]
\centering{
	\subfloat[Access to the vehicle (responses in \%).]{
		\includegraphics[trim = 2cm 3.5cm 2cm 3.5cm, clip, scale=0.25]{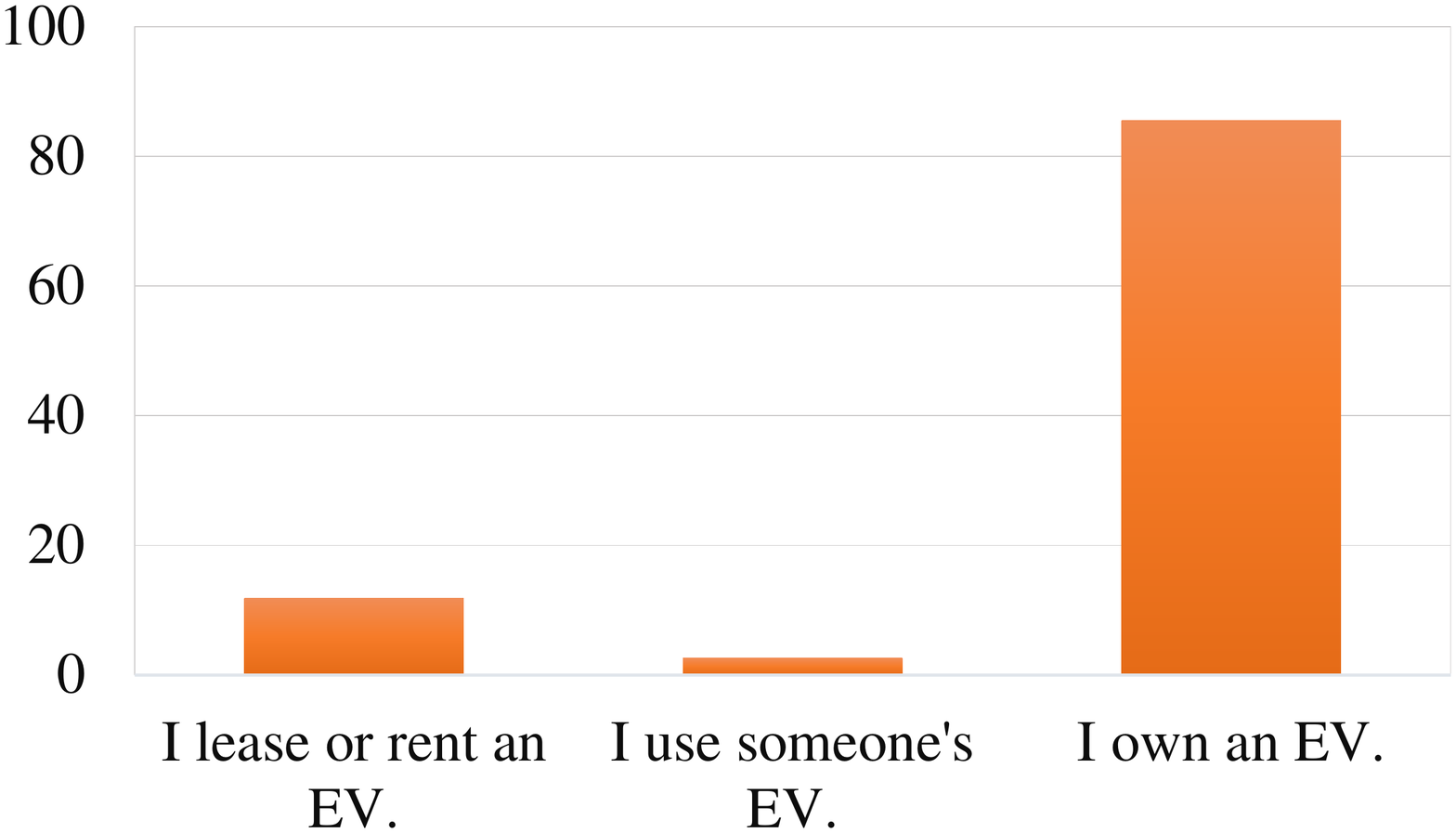}
		\label{fig:EVownershipDistribution}
	}
	\subfloat[Make of the vehicle (responses in \%).]{
		\includegraphics[trim = 2cm 4cm 2cm 3.5cm, clip, scale=0.35]{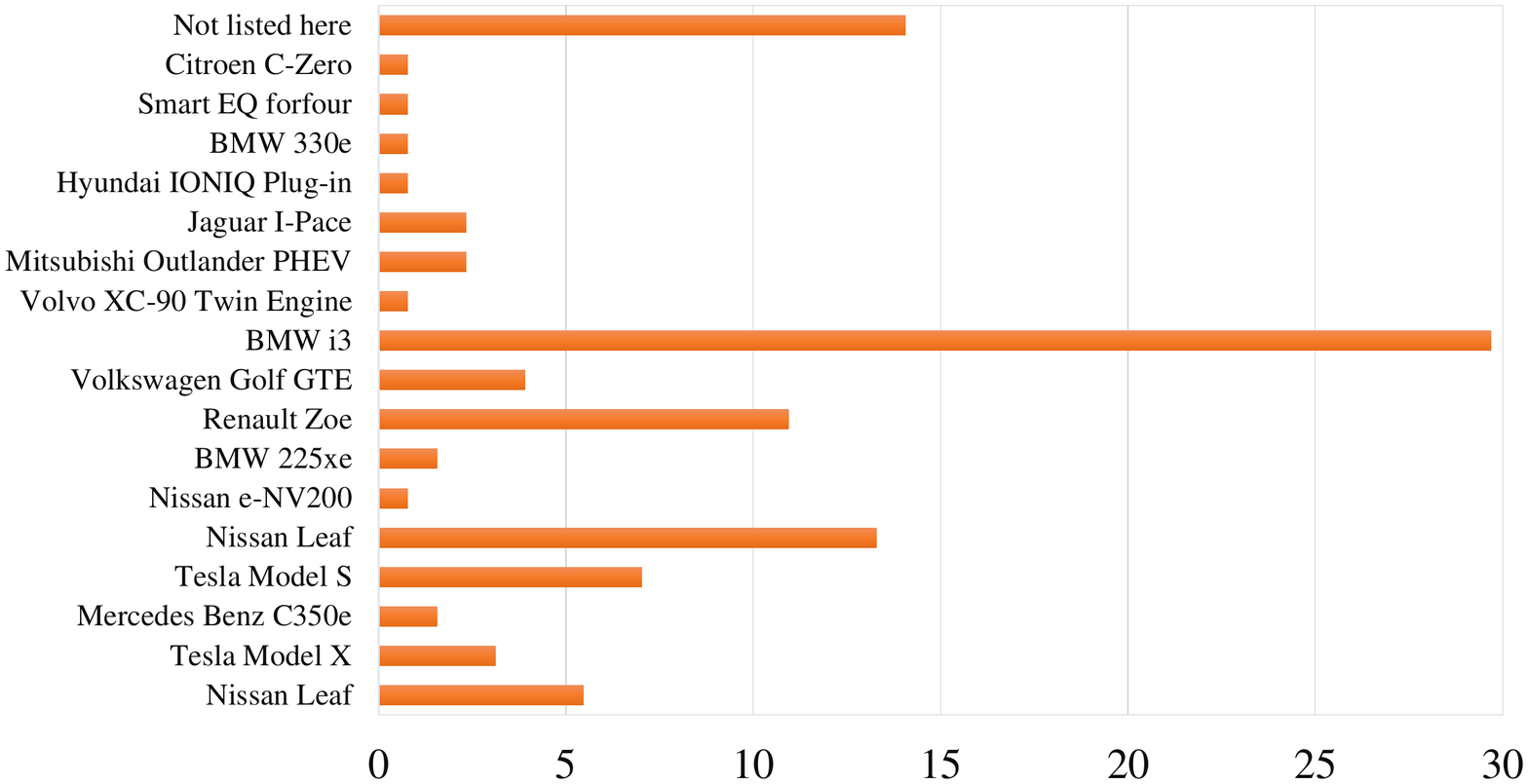}
		\label{fig:EVmakeDistribution}
	}
}
\caption{EV characteristics.}
\label{fig:aboutVehicle}
\end{figure*}

We use the vehicle as a discriminator of EV usage in our study. We specifically focus on two characteristics: 1) ownership, \textit{i.e.,} how the vehicle is accessed (owned, shared, or rented) and 2) make. While ownership can serve as an indicator of how the vehicle is used over time (\textit{i.e.,} successive drives and hence charge from different drivers), the make is needed to associate patterns of mobility with different energy profiles (the make affects the energy consumption). Figure \ref{fig:EVownershipDistribution} shows the characteristics of the surveyed drivers in terms of EV ownership. A majority of the sampled respondents declared to own an electric vehicle (more than 80\%). Figure \ref{fig:EVmakeDistribution} provides insights onto the types of EV driven in Greater London at the time of the survey.

\subsubsection{Travel habits characteristics}

\begin{figure*}[!t]
\centering{
	\subfloat[Daily trip duration (responses in \%).]{
		\includegraphics[trim = 2cm 6cm 2cm 5cm, clip, scale=0.30]{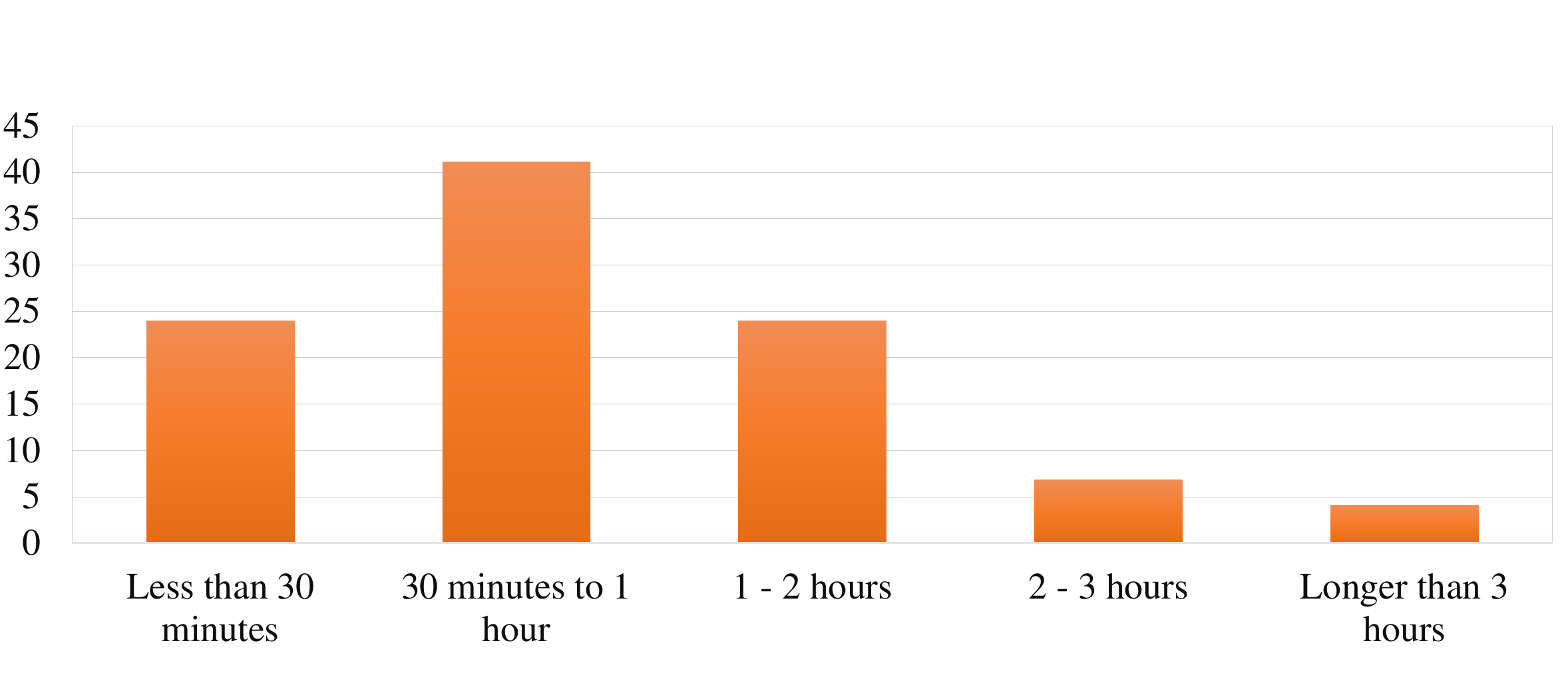}
		\label{fig:dailyTripDistribution}
	}\\
	\subfloat[EV usage purpose and frequency (responses in \%).]{
		\includegraphics[trim = 2cm 5.5cm 2cm 5cm, clip, scale=0.35]{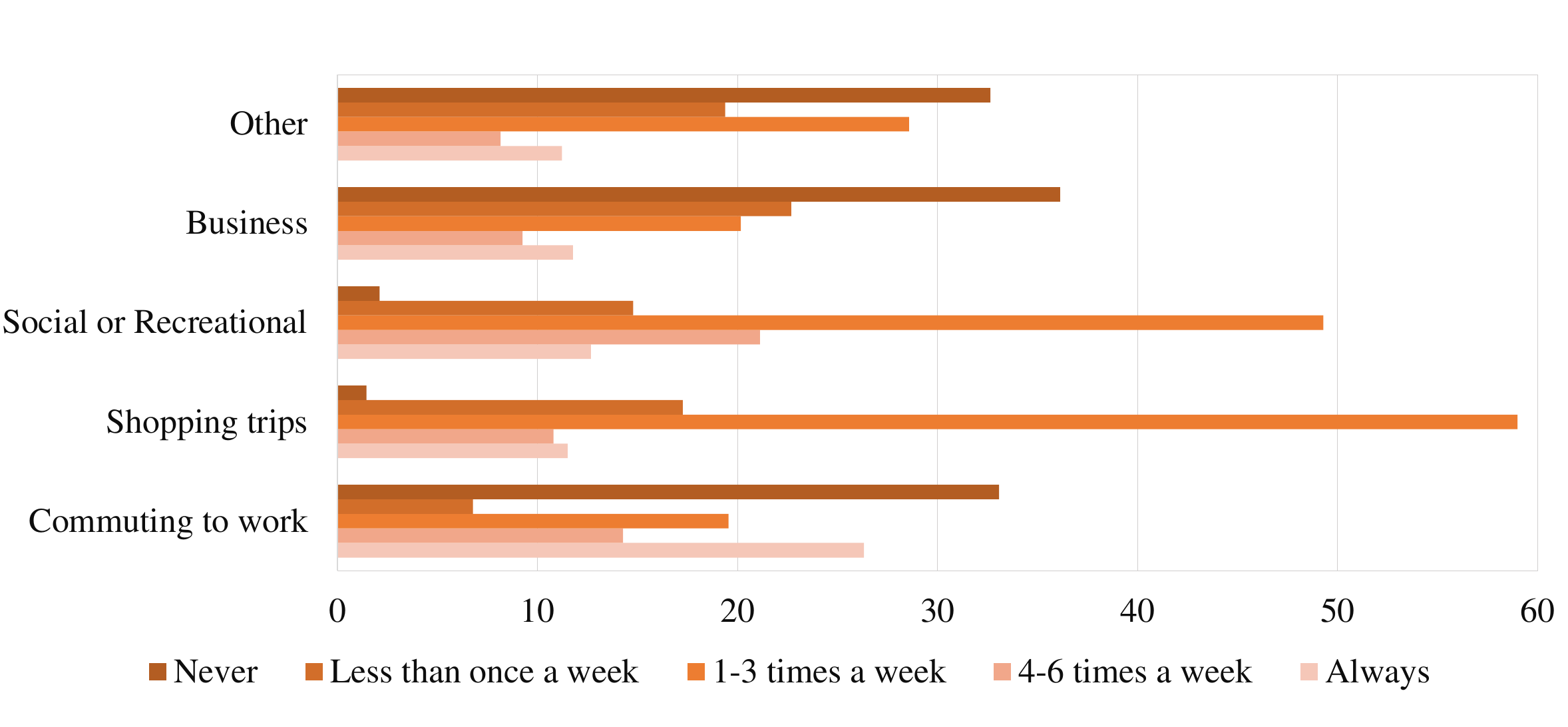}
		\label{fig:EVpurposeUsageDistribution}
	}\\
	\subfloat[EV usage purpose and trip duration (responses in \%).]{
		\includegraphics[trim = 2cm 5.5cm 2cm 5cm, clip, scale=0.35]{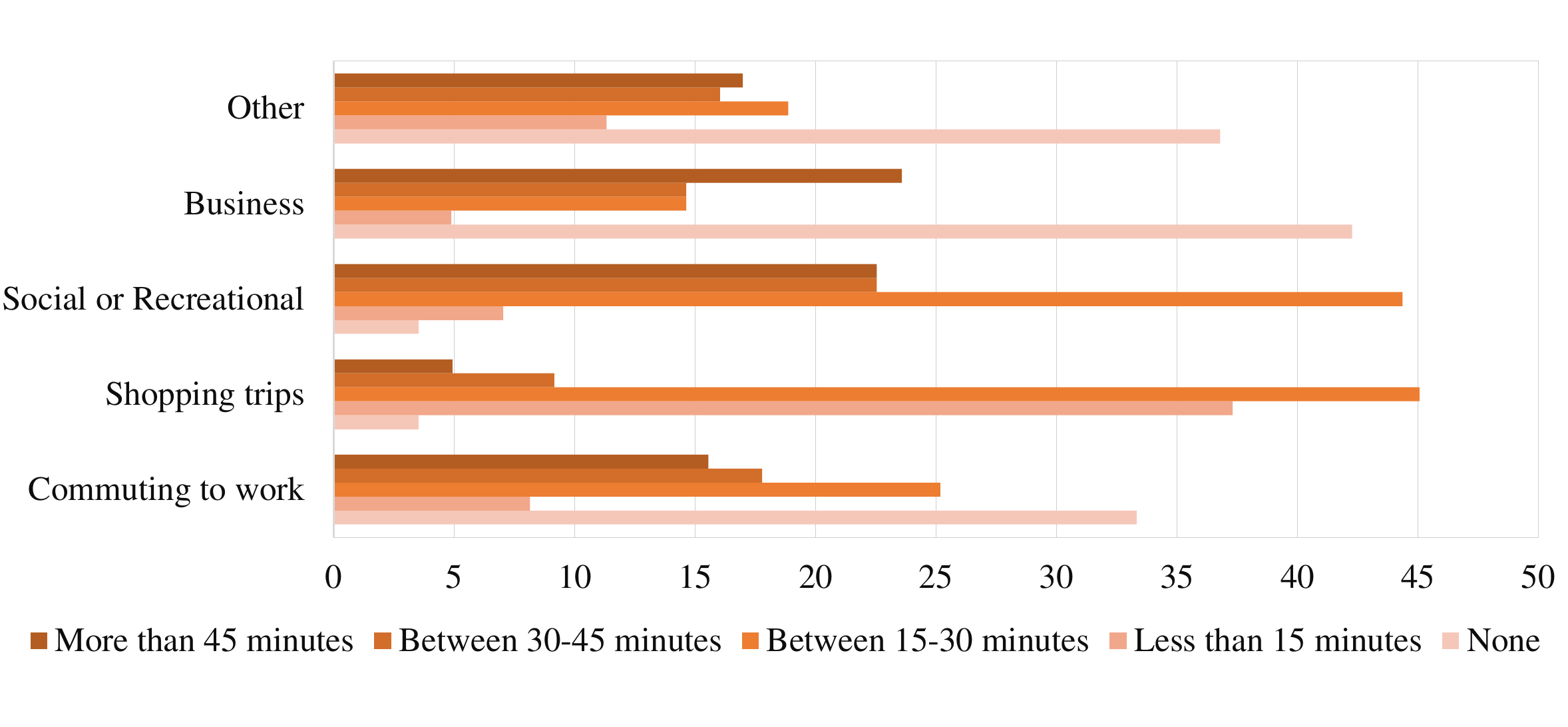}
		\label{fig:EVpurposeUsageTripDurationDistribution}
	}
}
\caption{Trip characteristics.}
\label{fig:aboutTravel}
\end{figure*}

The need for refueling stations (charging in the case of EV) depends on how users travel. The assessment of travel habits can thus provide important insights onto where infrastructure needs are likely to be predominant (in terms of the location of charging stations and their availability). We characterise travel habits through three attributes associated with mobility in an urban environment: 1) duration of daily trips using an EV, 2) frequency of using an EV as a function of different purposes, and 3) average trip duration in the context of each purpose. The distribution of the results is shown in Figure \ref{fig:aboutTravel}.

As shown in Figure \ref{fig:dailyTripDistribution}, the duration of the large majority of daily trips is below $2$ hours. As a comparison, the average commute times in London is estimated to be around 1h19\footnote{\url{https://www.tuc.org.uk/news/annual-commuting-time-21-hours-\\compared-decade-ago-finds-tuc}}. More than $40\%$ of the respondents declared making daily trips between 30 minutes to 1 hour. When asked about driving purpose(s), respondents indicated a diversity of reasons (Figure \ref{fig:EVpurposeUsageDistribution}), with a predominance in terms of frequency, for social/recreational trips and shopping. These purposes also correspond to the longest declared journeys (Figure \ref{fig:EVpurposeUsageTripDurationDistribution}).

\subsubsection{Charging habits characteristics}

The characteristics of charging habits were qualified in terms of accessibility and management of the charge.

\begin{figure*}[!t]
\centering{
	\subfloat[Home charger ownership (responses in \%).]{
		\includegraphics[trim = 2cm 6.5cm 2cm 6cm, clip, scale=0.32]{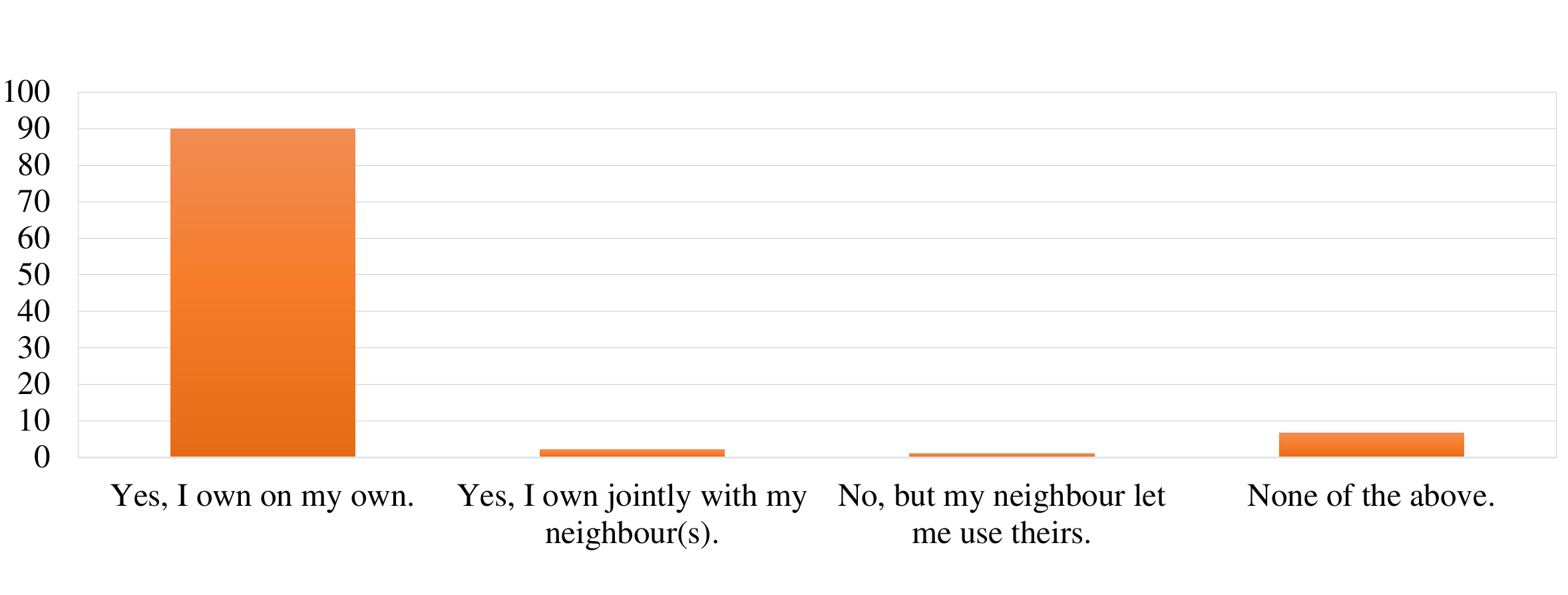}
		\label{fig:chargerOwnershipDistribution}
	}
	\subfloat[Home charger type (responses in \%).]{
		\includegraphics[trim = 2cm 6.5cm 2cm 6cm, clip, scale=0.32]{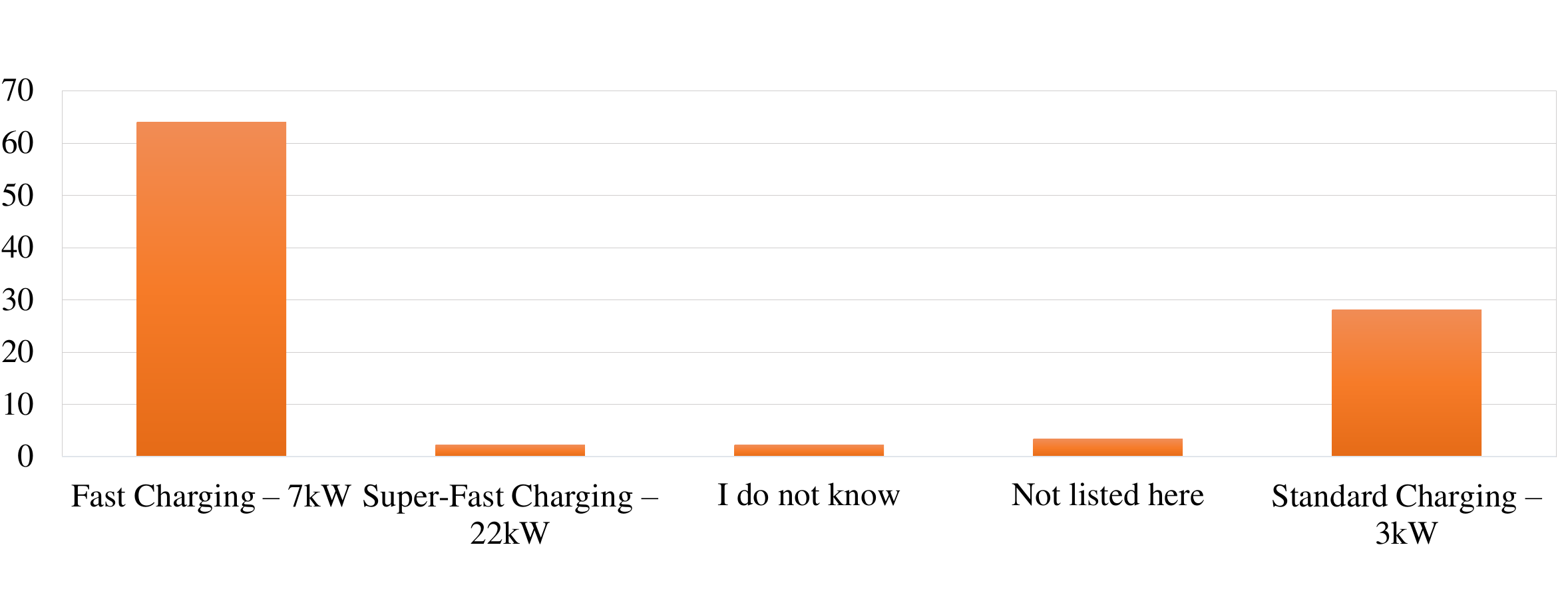}
		\label{fig:ownedTypedChargerDistribution}
	}
}
\vspace{-2mm}
\caption{Home charger access.}
\label{fig:aboutChargerOwnership}
\end{figure*}

\textbf{Accessibility} The first set of questions focuses on the accessibility to charging stations, which is determining to the experience of EV drivers can have. The results in Figure \ref{fig:chargerOwnershipDistribution} show the percentage of respondents with access to private charging. More than $90\%$ of the surveyed drivers declared owning a charger at home. A small percentage indicated sharing access to home charger with a neighbour. The type of charger owned is depicted in Figure \ref{fig:ownedTypedChargerDistribution}, with $65\%$ of the respondents owning a 7kW fast charger. 

\begin{figure}[hbt!]
\centering
\includegraphics[trim = 2cm 5cm 2cm 5cm, clip, scale=0.375]{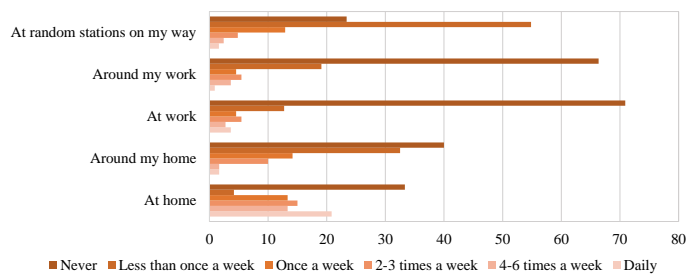}  
\vspace{-5mm}
\caption{Popularity of charging locations (responses in \%).}
\label{fig:aboutChargeLocationFrequency}
\end{figure}

Figure \ref{fig:aboutChargeLocationFrequency} focuses on the popularity of charging location, more specifically home and work charging. The results show where respondents declared to mostly charge their vehicle over a period of a week. Home charging is the method used the most on a daily basis.   

\begin{figure*}[!t]
\centering{
	\subfloat[Public charger usage (responses in \%).]{
		\includegraphics[trim = 2cm 6.5cm 2cm 6cm, clip, scale=0.35]{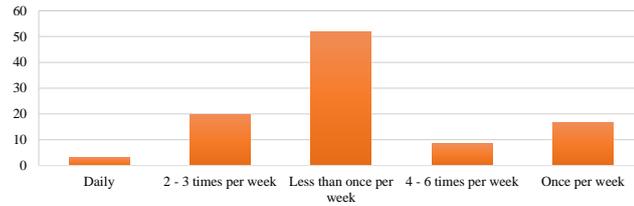}
		\label{fig:frequencyPublicChargingUsageDistribution}
	}\\
	\subfloat[Criticality of factors affecting the use of public stations (responses in \%).]{
		\includegraphics[trim = 2cm 5cm 2cm 5cm, clip, scale=0.35]{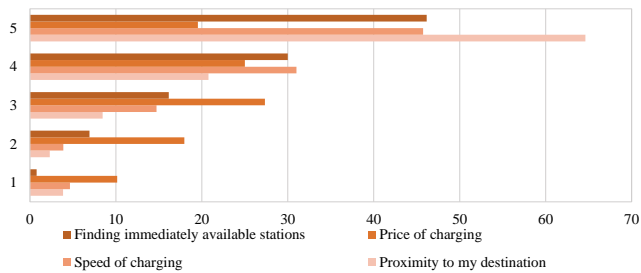}
		\label{fig:factorsUsingPublicChargingDistribution}
	}
}
\vspace{-2mm}
\caption{Access to public charging stations.}
\label{fig:aboutPublicChargingUsage}
\end{figure*}

The next set of results in Figure \ref{fig:aboutPublicChargingUsage} focuses on the access to public charging networks. Figure \ref{fig:frequencyPublicChargingUsageDistribution} shows how often respondents declared to make use of a public charging station in a week. Figure \ref{fig:factorsUsingPublicChargingDistribution} provides insights regarding how critical a factor is for a driver to use public charging stations (from 1 not critical to 5 highly critical). The proximity of stations to the destination of the journey is cited as a highly critical factor by $65\%$ of the respondents. Availability and speed of charging are listed as the second two most important factors.

\begin{figure*}[!t]
\centering{
	\subfloat[Way to locate public charging stations (responses in \%).]{
		\includegraphics[trim = 2cm 6.5cm 2cm 6cm, clip, scale=0.60]{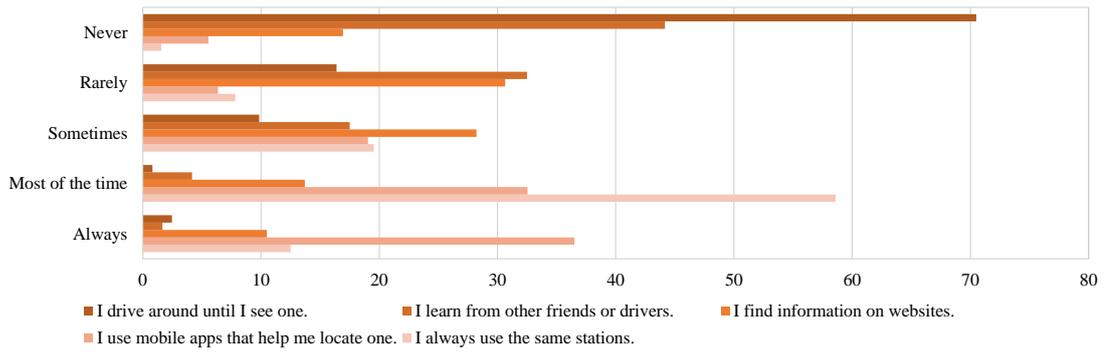}
		\label{fig:publicChargingDiscoveryMethodDistribution}
	}\\
	\subfloat[Proximity between the selected charging station and the destination (responses in \%).]{
		\includegraphics[trim = 2cm 6.5cm 2cm 6cm, clip, scale=0.50]{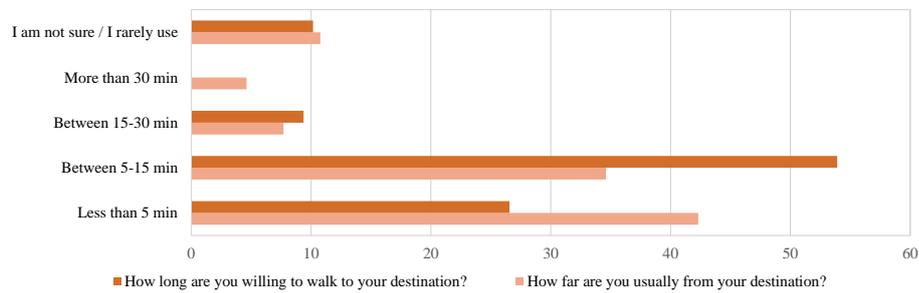}
		\label{fig:considerationsUsingPublicChargingDistribution}
	}
}
\caption{Accessibility to public charging stations.}
\label{fig:aboutPublicChargingAccessibility}
\end{figure*}

The access to public charging stations depends on whether drivers are informed about the location of stations. Figure \ref{fig:publicChargingDiscoveryMethodDistribution} shows how respondents are informed about the location of public charging stations. For convenience, drivers tend to use the same stations ($58\%$ of the respondents declared to use the same station most of the time). Mobile apps are reported as the primary source of information by more than $35\%$ of the respondents. Figure \ref{fig:considerationsUsingPublicChargingDistribution} focuses on the importance of proximity for charging. The majority of the respondents declared to use stations that are located less than $15$ minutes walk to their destination to charge their vehicle, and are in general reluctant to walk longer. 

\begin{figure*}[!t]
\centering{
	\subfloat[Time of charge (responses in \%).]{
		\includegraphics[trim = 2cm 5cm 2cm 5cm, clip, scale=0.30]{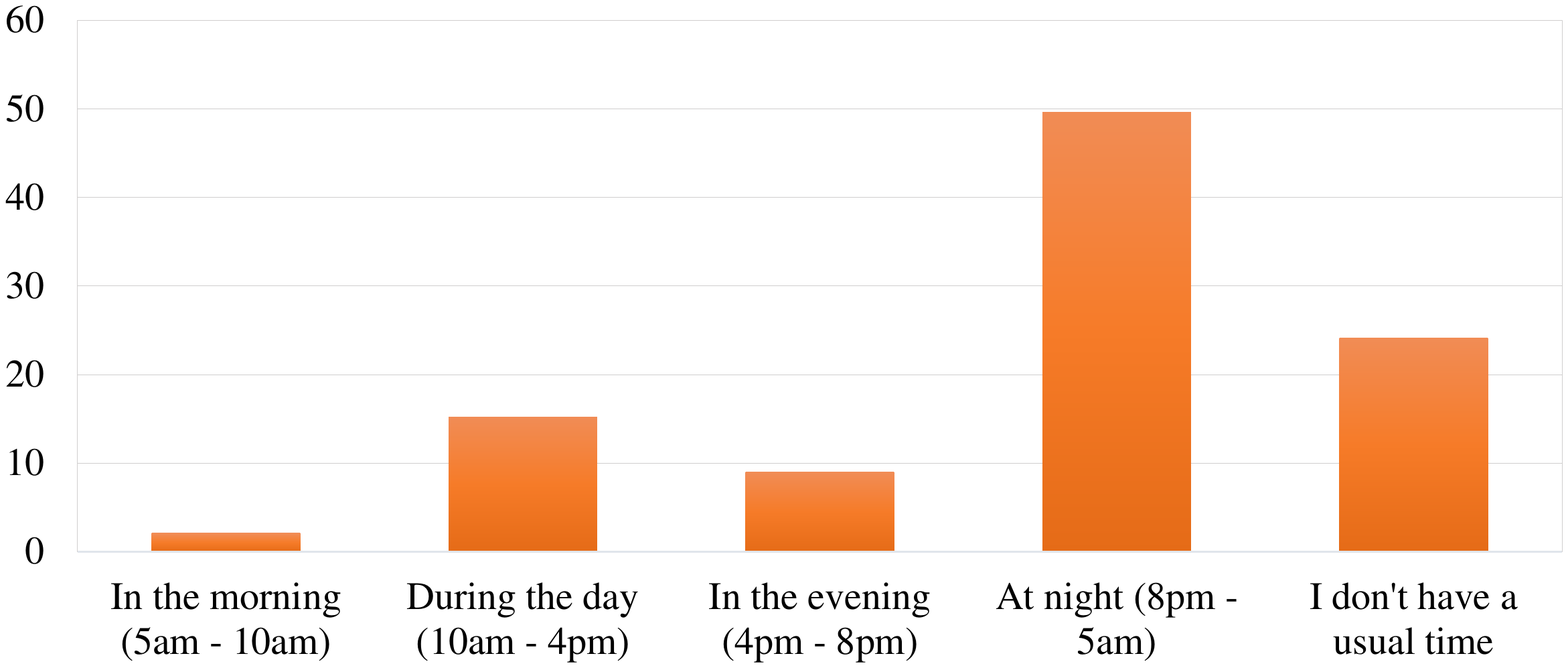}
		\label{fig:timeOfCharge}
	}
	\subfloat[Frequency vehicle charging (responses in \%).]{
		\includegraphics[trim = 2cm 6.5cm 2cm 6cm, clip, scale=0.30]{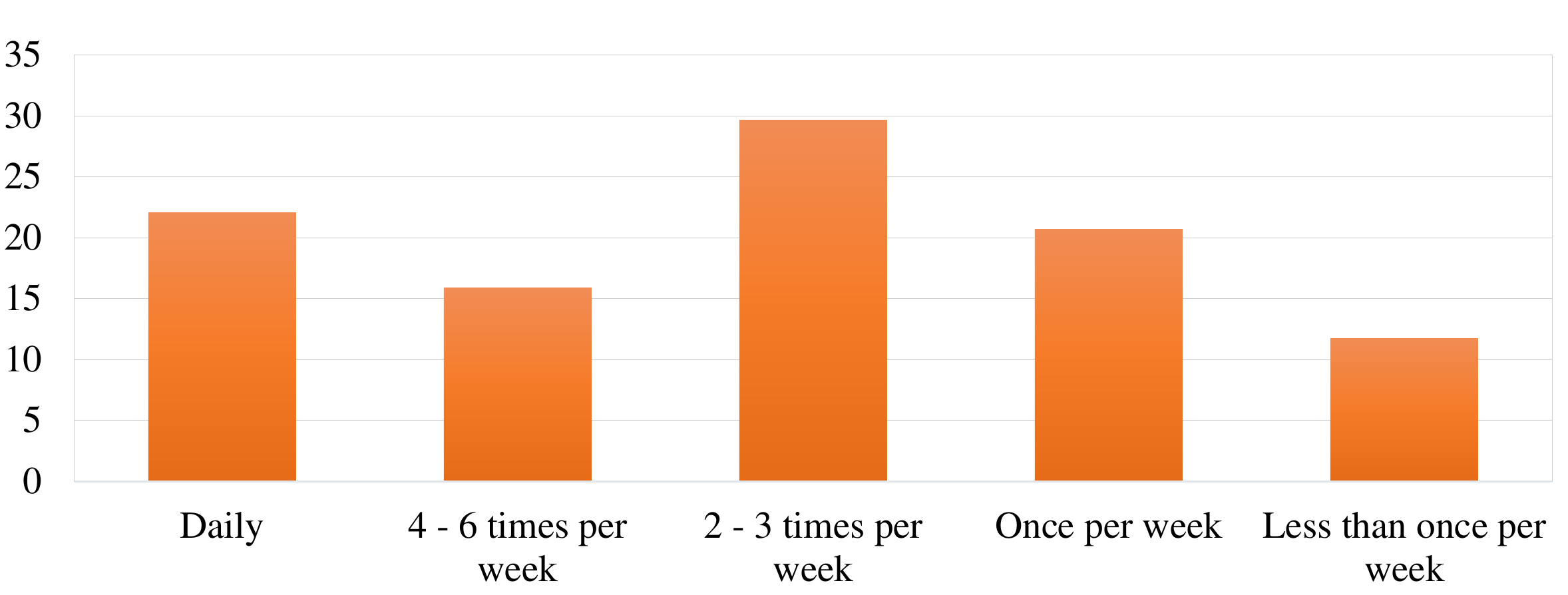}
		\label{fig:frequencyVehicleChargingDistribution}
	}
}
\vspace{-3mm}
\caption{Charging time and frequency.}
\label{fig:aboutChargingPeriod}
\end{figure*}

\textbf{Charge management behaviour} The second set of questions focuses on the behaviour of drivers with respect to the management of the charge of their vehicle. Figure \ref{fig:aboutChargingPeriod} shows when respondents tend to charge and how often they recharge their vehicle on a weekly basis. In line with the fact that home charging is identified as the predominant charging method in the collected responses, almost $50\%$ of the surveyed drivers reported to mostly performed charging at night (Figure \ref{fig:timeOfCharge}). The second most popular answer is the absence of a usual time. The frequency of recharge is shown in Figure \ref{fig:frequencyVehicleChargingDistribution}, with the frequency of 2-3 times per week selected by the largest number of respondents ($30\%$). Interestingly, the daily and once-per-week frequencies were choosen in almost the same proportions.  

\begin{figure}[hbt!]
\centering
\includegraphics[trim = 2cm 5cm 2cm 6cm, clip, scale=0.35]{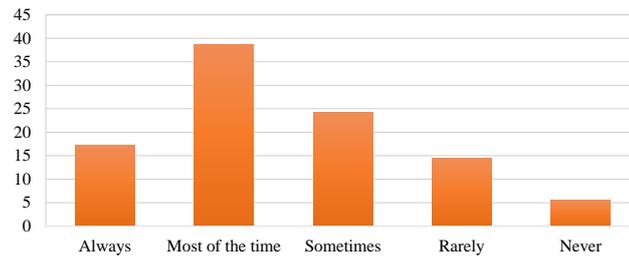}  
\vspace{-3mm}
\caption{Recharge planning (responses in \%).}
\label{fig:aboutChargePlanning}
\end{figure}

Figure \ref{fig:aboutChargePlanning} shows how respondents plan the recharge of their vehicle. While most of the respondents indicated to plan the recharge most the time, responses are diverse with some drivers never planning and some drivers always planning.  

\begin{figure*}[!t]
\centering{
	\subfloat[Criteria to recharge (responses in \%).]{
		\includegraphics[trim = 2cm 5cm 2cm 5cm, clip, scale=0.30]{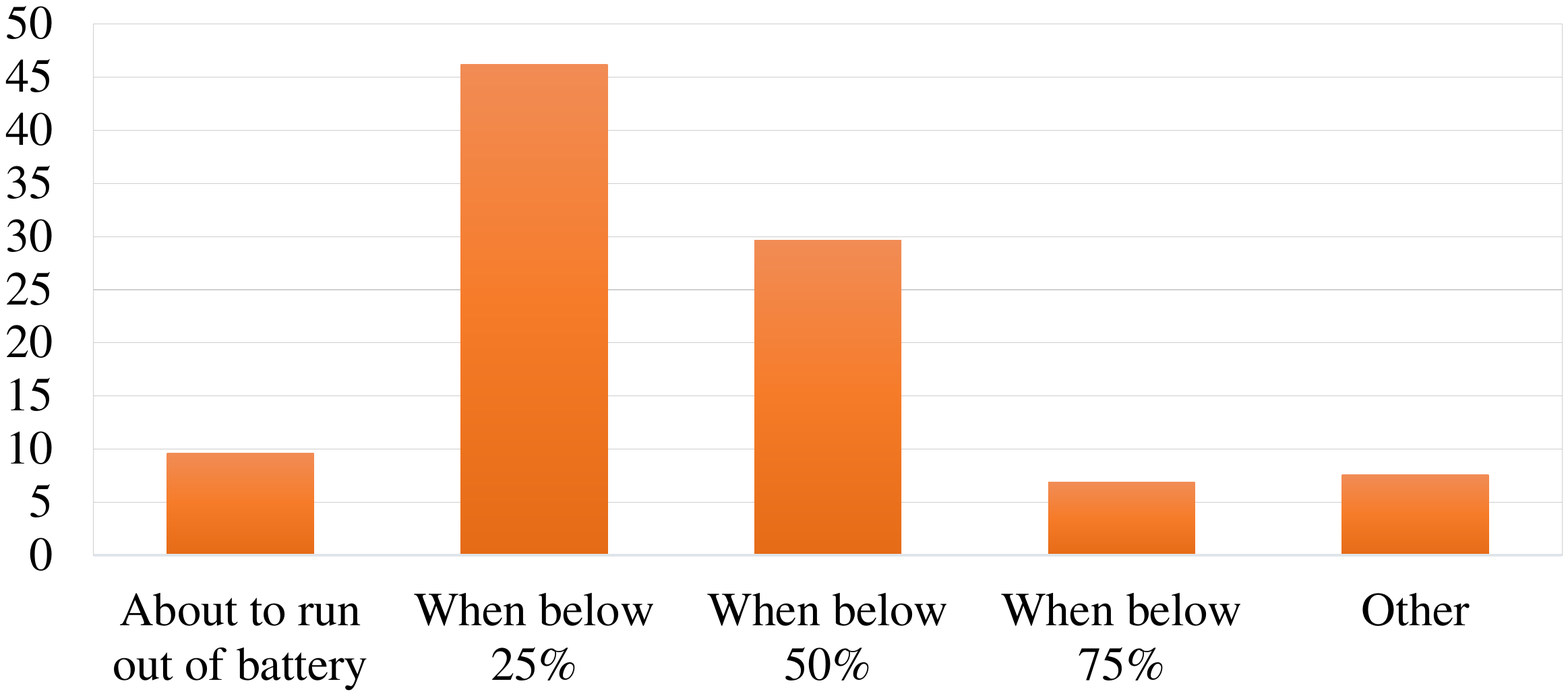}
		\label{fig:batteryLevelStartChargingDistribution}
	}
	\subfloat[Criteria to stop a charge (responses in \%).]{
		\includegraphics[trim = 2cm 6.5cm 2cm 6cm, clip, scale=0.35]{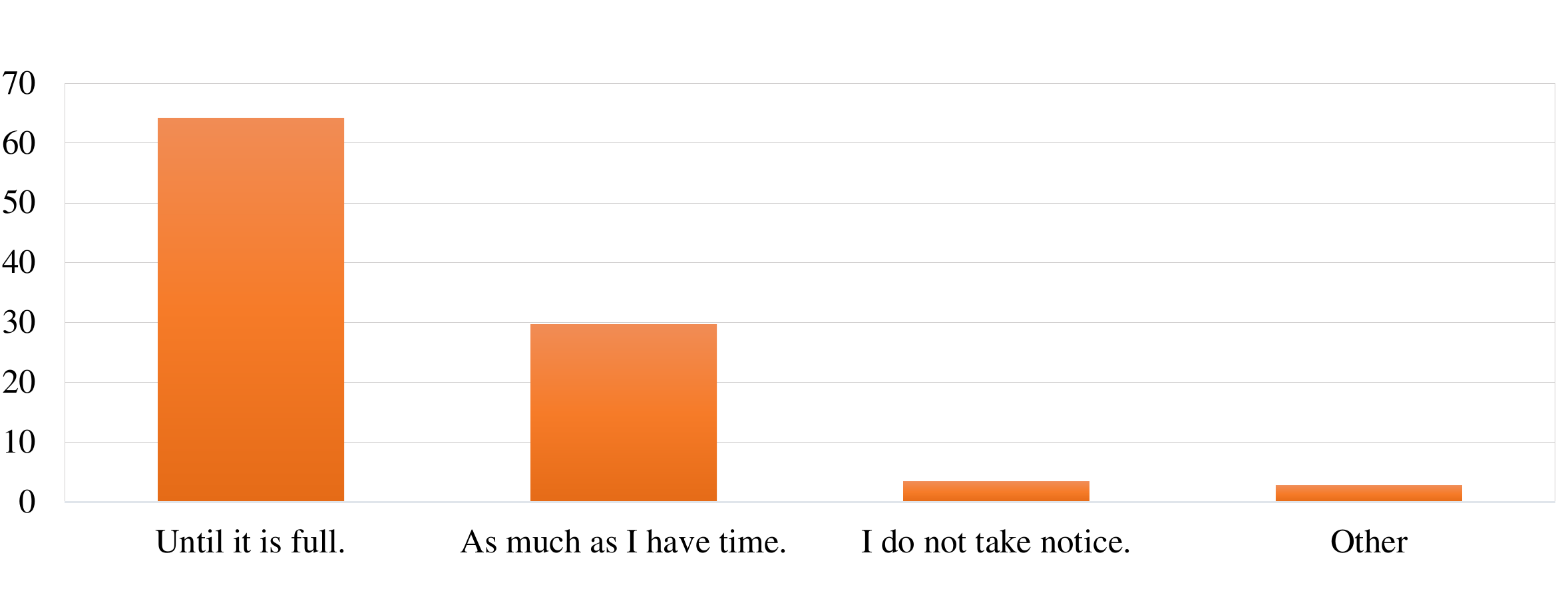}
		\label{fig:upToWhichLevelKeepChargingDistribution}
	}
}
\caption{Charge management.}
\label{fig:aboutChargeManagement}
\end{figure*}

Figure \ref{fig:aboutChargeManagement} shows the incidence of the battery level on the decision to 1) recharge and 2) stop a charge. Drivers exhibit different behaviours with respect to when they decide to recharge their vehicle, from highly precautions-type of drivers recharging when the battery level falls under 75\% to adventurous-type of drivers waiting for the battery to run out before recharging. Most of the respondents indicated to wait for the battery level to fall below $50\%$ before recharging, with the largest number of replies obtained for a battery level below $25\%$. Figure \ref{fig:upToWhichLevelKeepChargingDistribution} focuses on the criteria respondents tend to use for deciding when to stop a charge. The majority of the respondents indicated to wait for the full charge to stop. This can be put into perspective with the fact that most respondents use home charger and charge at nights. The second most popular answer is as much time permits, which applies well in case the charge is at intermediate stops on a journey.

\begin{figure*}[!t]
\centering{
	\subfloat[Monthly budget for charging (responses in \%).]{
		\includegraphics[trim = 2cm 6cm 2cm 5cm, clip, scale=0.30]{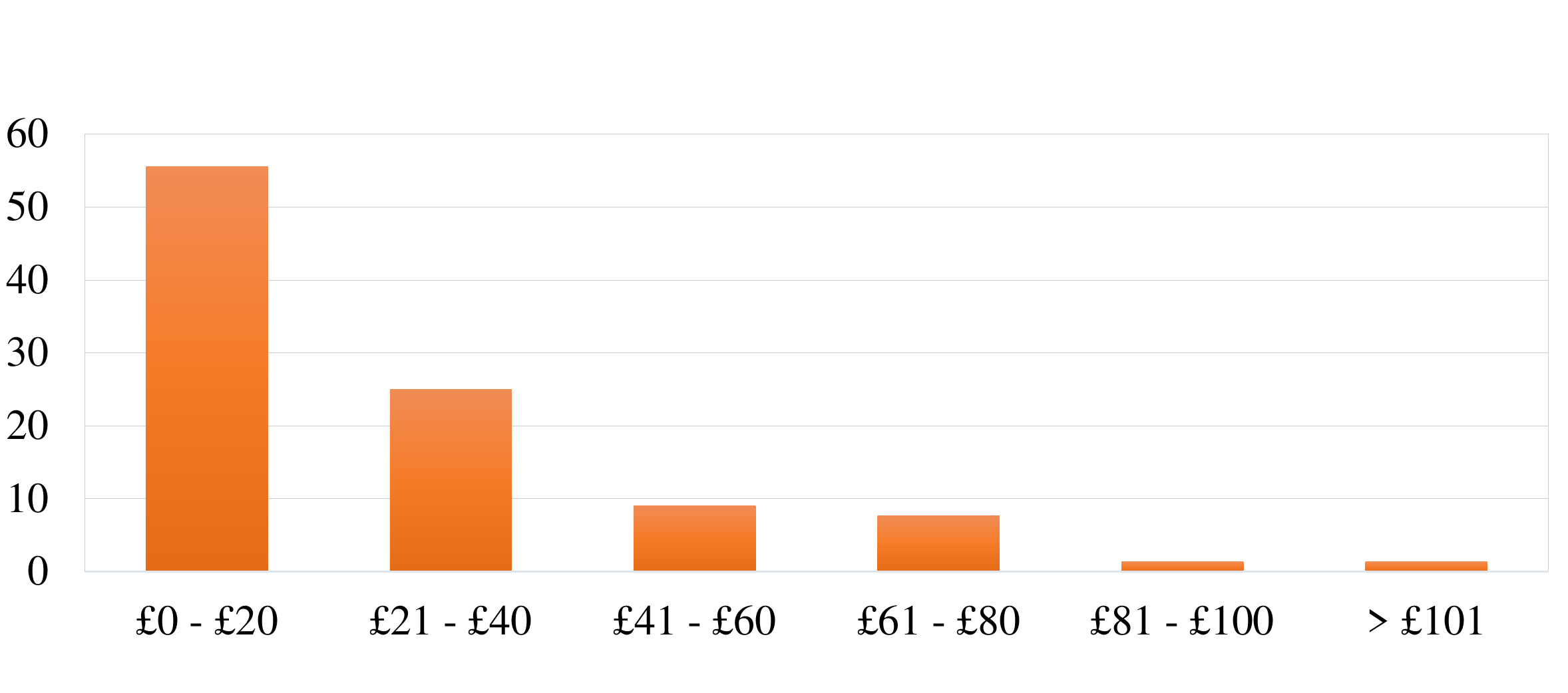}
		\label{fig:monthlyBudgetChargingDistribution}
	}
	\subfloat[Agreement with the statement "the charging budget is significant" (responses in \%)]{
		\includegraphics[trim = 2cm 6cm 2cm 5cm, clip, scale=0.30]{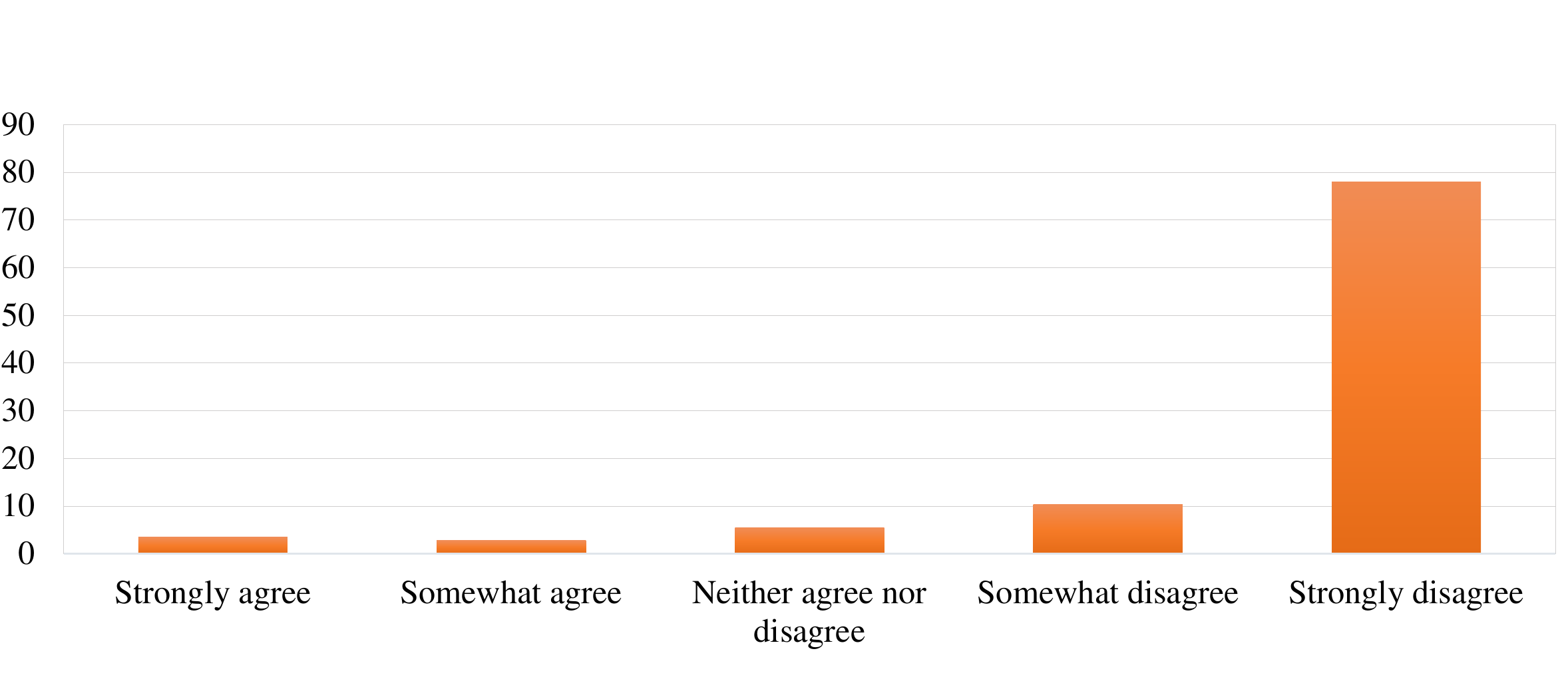}
		\label{fig:chargingBudgetStatementAgreementDistribution}
	}
}
\caption{Charging costs.}
\label{fig:aboutChargeBudgetManagement}
\end{figure*}

The results depicted in Figure \ref{fig:aboutChargeBudgetManagement} concern charging costs. Figure \ref{fig:monthlyBudgetChargingDistribution} shows how much respondents declared to spend on charging per month. Charging costs do not amount to more than £20 a month in the large majority of cases\footnote{More detailed questions would be needed to understand whether responses indicating a monthly budget higher than £80 could be classified as outliers.}. Figure \ref{fig:chargingBudgetStatementAgreementDistribution} shows how respondents perceive the budget they spend on charging. For most of them, charging is not considered to constitute a significant budget of using an EV. 

\subsubsection{About charging service preferences}

The type of services offered on the public charging network in terms of reserved time slot, preferential charge rate, \textit{etc.}, can have an influence on how users access the public charging stations.

\begin{figure}[hbt!]
\centering
\includegraphics[trim = 2cm 6cm 2cm 6cm, clip, scale=0.30]{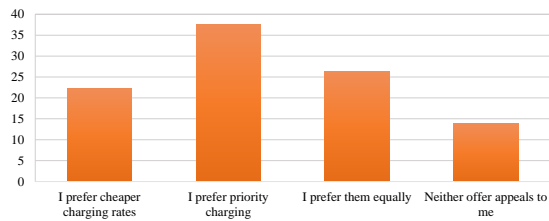}  
\vspace{-2mm}
\caption{Responsiveness to charging offers (responses in \%).}
\label{fig:aboutPreferredOffer}
\end{figure}

Figure \ref{fig:aboutPreferredOffer} shows the receptiveness of respondents to three charging offers: 1) priority charging, \textit{i.e.,} getting priority access to a charging station, 2) reduced charge rate, and 3) a combination of priority charging and reduced charge rate. The largest number of respondents selected priority charging as their preferred offer ($37\%$). Between the reduced charging rate option and the combined option, a larger number of drivers declared to prefer the latter. It is to be noted that almost $15\%$ of the respondents indicated to neither be interested in getting priority access nor being offered reduced charging rates. 

\begin{figure*}[!t]
\centering{
	\subfloat[Deporting the charge in time (responses in \%).]{
		\includegraphics[trim = 2cm 6cm 2cm 6cm, clip, scale=0.30]{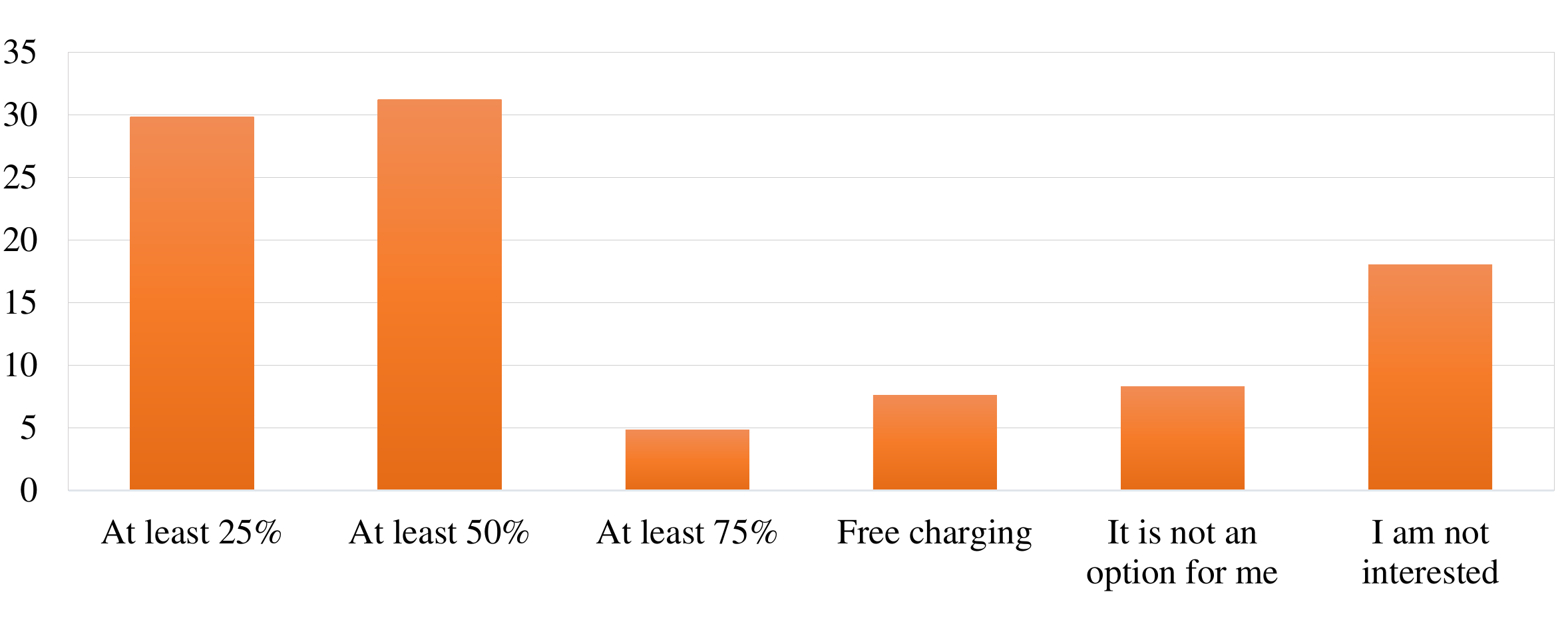}
		\label{fig:acceptanceOfferedReducedChargingRateForDelayingDistribution}
	}
	\subfloat[Deporting the charge in space (responses in \%).]{
		\includegraphics[trim = 2cm 6cm 2cm 6cm, clip, scale=0.30]{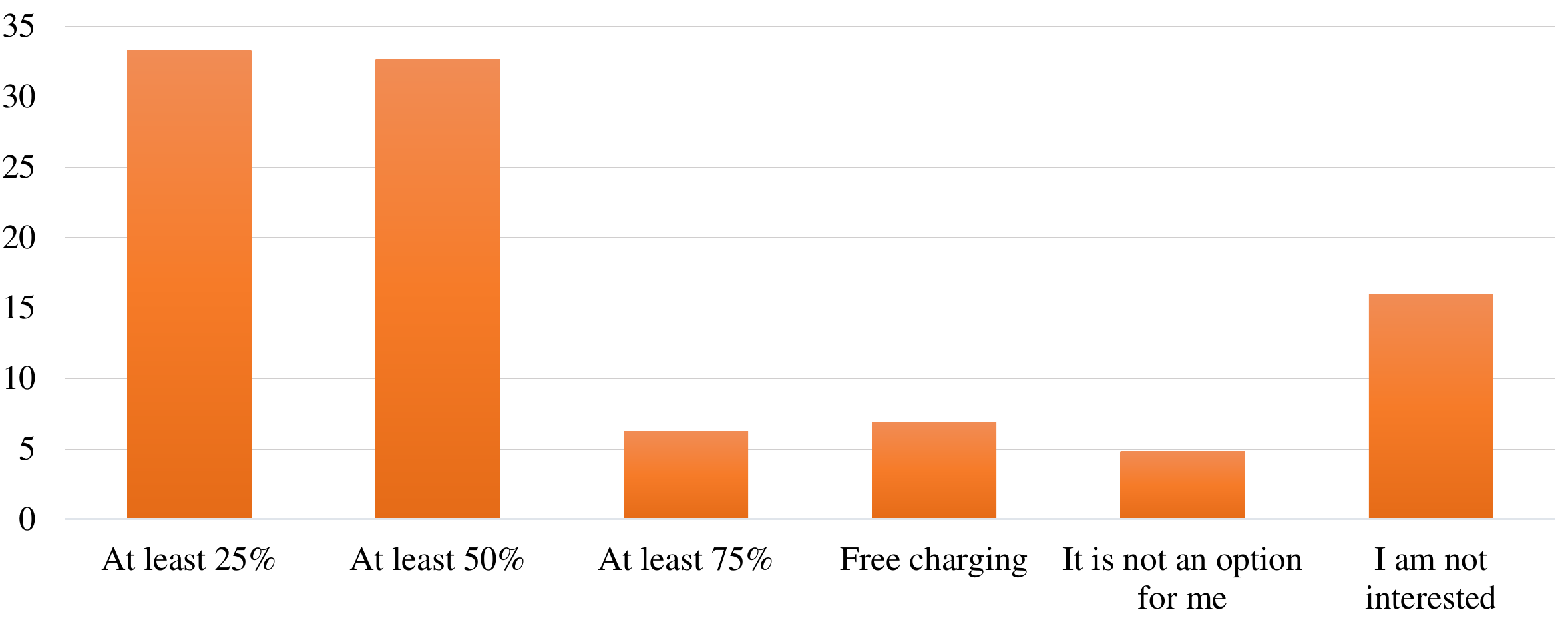}
		\label{fig:acceptanceOfferedReducedChargingRateForWalkingDistribution}
	}
}
\caption{Responsiveness to a reduction of the tariff of a charge.}
\label{fig:aboutOfferAcceptanceConditionsMoney}
\end{figure*}

Figure \ref{fig:aboutOfferAcceptanceConditionsMoney} focuses on the responsiveness of drivers to a reduction in the tariff of a charge in exchange to subscribing to a charge deporting scheme, namely 1) deporting the charge in time, \textit{i.e.,} the driver is asked to wait to charge (Figure \ref{fig:acceptanceOfferedReducedChargingRateForDelayingDistribution}), and 2) deporting the charge in space, \textit{i.e.,} the driver is redirected to a charging station that is located further away from the destination (Figure \ref{fig:acceptanceOfferedReducedChargingRateForDelayingDistribution}). The tariff's reduction under which respondents would consider subscribing to such a scheme is similar for both deporting options: a same proportion of around $30\%$ of the drivers would be responsive to a reduction of either $25\%$ and $50\%$. 

\begin{figure}[hbt!]
\centering
\includegraphics[trim = 2cm 5cm 2cm 5cm, clip, scale=0.35]{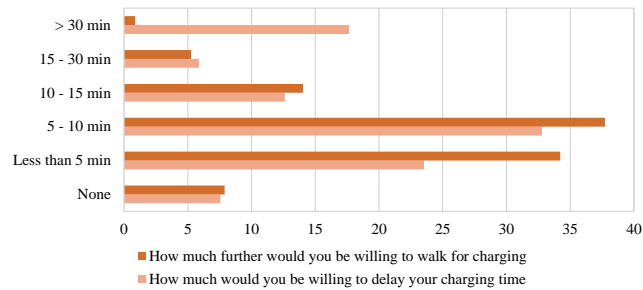}  
\vspace{-2mm}
\caption{Willingness to spend more time on the charge (responses in \%).}
\label{fig:aboutOfferAcceptanceConditionsTimeLoss}
\end{figure}

Figure \ref{fig:aboutOfferAcceptanceConditionsTimeLoss} focuses on the willingness of drivers to spend more time on their charge by either accepting to delay it or being redirected to a station that is located further away from their destination. The results show that the time acceptance is in the order of $5$ to $10$ minutes, beyond which respondents tend to be less responsive to the perspective of spending more time.     

\begin{figure}[hbt!]
\centering
\includegraphics[trim = 2cm 6cm 2cm 6cm, clip, scale=0.35]{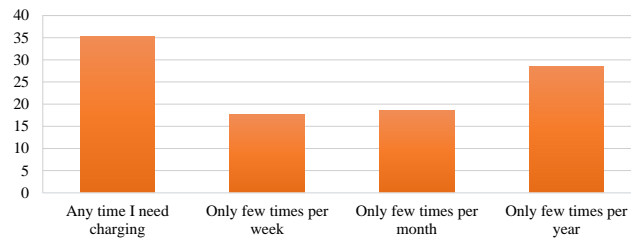}  
\vspace{-2mm}
\caption{Responsiveness to the recurrence of an offer (responses in \%).}
\label{fig:aboutFrequencyOfferAcceptance}
\end{figure}

Figure \ref{fig:aboutFrequencyOfferAcceptance} shows how often drivers would be willing to accept service offers. While the largest number of respondents indicated that they would be interested in getting such offers any time they need to charge ($35\%$), a non-negligible number of them declared not being interested in getting these offers often (almost $20\%$ only a few times per month and almost $30\%$ only a few times per year). 

\begin{figure*}[!t]
\centering{
	\subfloat[Deporting charging in time (responses in \%).]{
		\includegraphics[trim = 2cm 6cm 2cm 6cm, clip, scale=0.35]{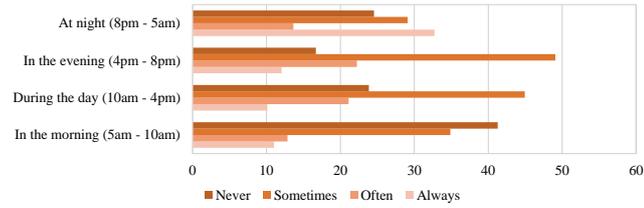}
		\label{fig:acceptanceOfferedReducedChargingRateForDelayingDistribution}
	}\\
	\subfloat[Deporting charging in space (responses in \%).]{
		\includegraphics[trim = 2cm 6cm 2cm 6cm, clip, scale=0.35]{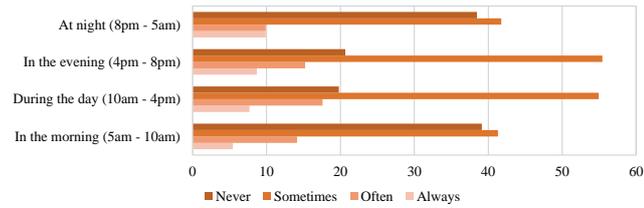}
		\label{fig:acceptanceOfferedReducedChargingRateForWalkingDistribution}
	}
}
\caption{Influence of the time of the day on drivers' responsiveness to deporting charging schemes.}
\label{fig:aboutInfluenceTimeOfDayAcceptanceOffers}
\end{figure*}

\begin{figure*}[!t]
\centering{
	\subfloat[Deporting charging in time (responses in \%).]{
		\includegraphics[trim = 2cm 6cm 2cm 6cm, clip, scale=0.35]{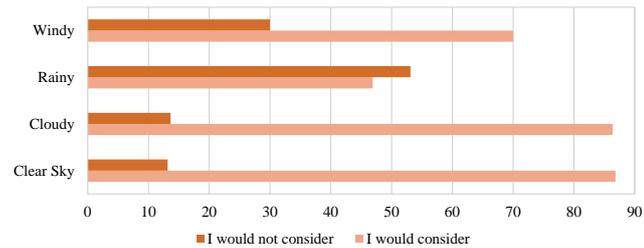}
		\label{fig:acceptanceOfferedReducedChargingRateForDelayingDistribution}
	}\\
	\subfloat[Deporting charging in space (responses in \%).]{
		\includegraphics[trim = 2cm 6cm 2cm 6cm, clip, scale=0.35]{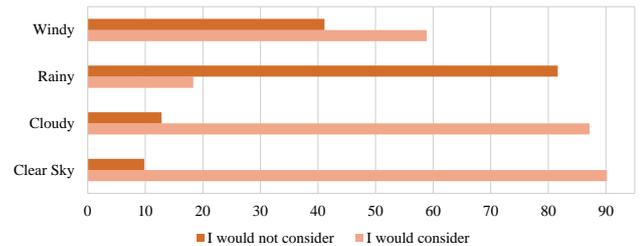}
		\label{fig:acceptanceOfferedReducedChargingRateForWalkingDistribution}
	}
}
\caption{Influence of the weather on drivers' responsiveness to deporting charging schemes.}
\label{fig:aboutInfluenceWeatherAcceptanceOffers}
\end{figure*}

Finally, Figure \ref{fig:aboutInfluenceTimeOfDayAcceptanceOffers} and Figure \ref{fig:aboutInfluenceWeatherAcceptanceOffers} focus on the influence of the time of the day and the weather, respectively, on the responsiveness of drivers to the deporting charging schemes. 

\subsubsection{Comments on the applicability of survey results}

As of March 2022, there are $768,329$ plug-in vehicles registered in the UK as opposed to $264,486$ in 2019 \cite{zapmapStat}. In terms of the public charging infrastructure, the number of connectors has increased from $20,964$ in 2019 to $29,664$ \cite{zapmapNetw}. The observations that can as such be extracted from the results of the survey we carried out need to be placed in their relevant context. In particular, the obtained sample is by its nature biased to the area (Greater London) where results were collected and the fact that the responses mostly came from early adopters of EVs. While the results may not be representative of the general situation in 2022, they provide a user-focused perspective onto a service that has been rapidly developing over the last three years. 

\subsection{Consultations with electric vehicle drivers}

We complemented the insights that were obtained through the survey by running consultations with EV drivers in Greater London. The objective of the consultations was to discuss the experience of drivers when using the public charging infrastructure. The participants insisted on five main aspects impairing their experience of public charging networks. 

They first reported that it was highly cumbersome to manage individual subscription to different charging network providers. In 2019, public charging stations in the Greater London region were operated by a dozen of providers, each requiring prospective users to subscribe to their service in order to be able to access the stations they operate. Given that providers do not have the same network coverage across the Greater London region, drivers are indirectly incited to subscribe charging services from different providers, each with its own specifics. In addition, some participants mentioned that the absence of charging network coverage in their neighborhood was a critical factor affecting their lack of motives for using the public charging network. They also indicated that on top of not necessarily having the option to use a close-by public station, existing stations were often not available or functioning properly for them to be considered as a reliable primary source of charging. The lack of a sufficient number of stations offering super-fast charging was also reported as hindering factor. Finally participants expressed that it was not always easy to find / locate stations and that existing applications had shortcomings when it comes to providing quality information regarding station's accessibility.  



\section{Impact of the e-mobility demand on the infrastructure}
\label{sect:demandModel}

Different components of the e-mobility demand need to be taken into account in order to study the impact of the demand on the public charging infrastructure. We developed a framework to guide the estimation of these components based on the Greater London use case. The framework operates at the macroscopic level. It aims at providing high-level insights on whether the public charging infrastructure is sufficient to satisfy the charging needs of the fleet of EVs in a pre-determined region.  

\subsection{Estimation of e-mobility demand components}

The framework is composed of three estimation modules: 

\begin{enumerate}
	\item the average daily energy needs of an EV;
	\item the overall daily charging needs of the fleet of EVs;
	\item the geographical distribution of the public charging needs.
\end{enumerate}

\begin{figure}[hbt!]
\centering
\includegraphics[trim = 0cm 4cm 0cm 0cm, clip, scale=0.55]{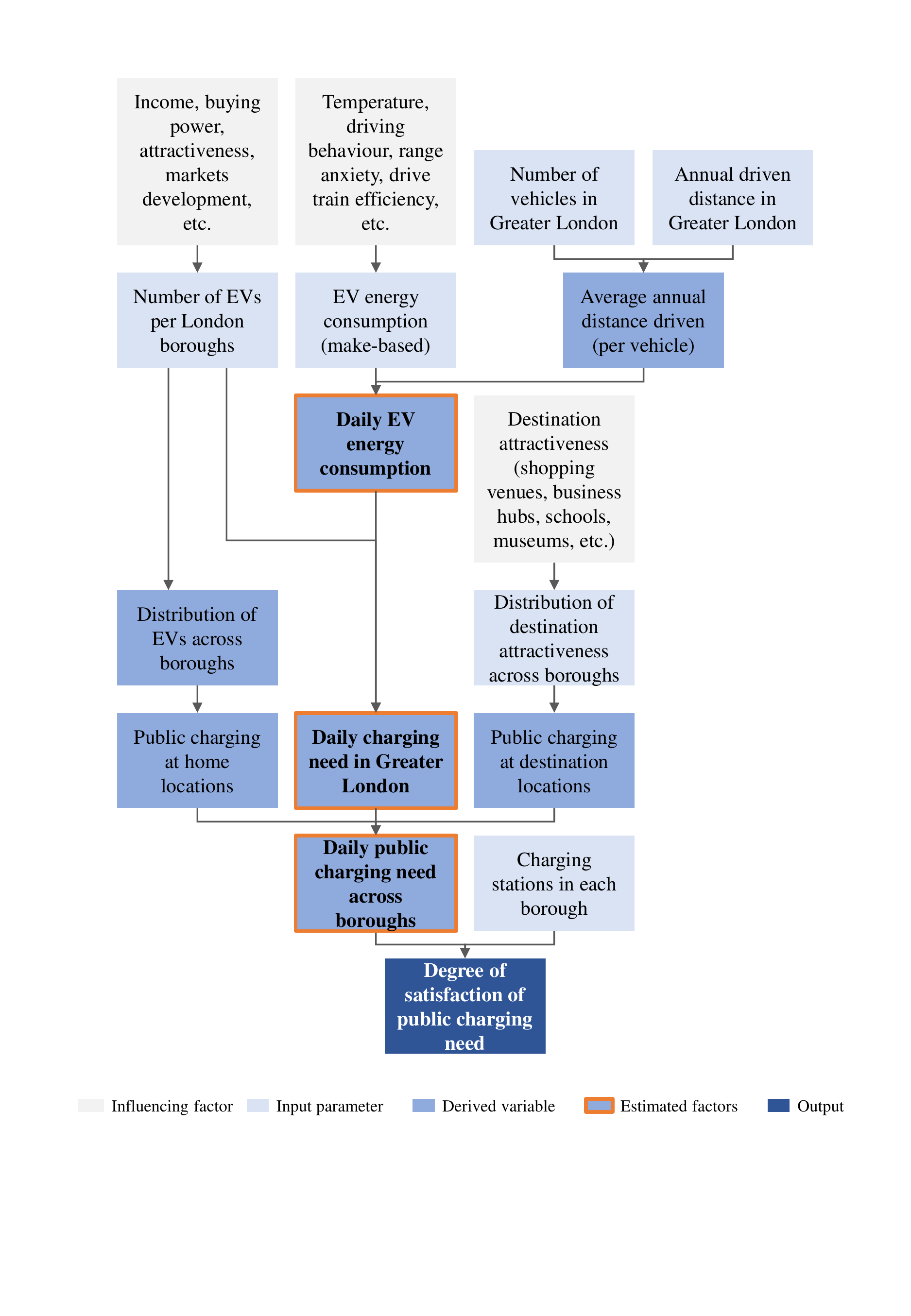}  
\vspace{-2mm}
\caption{Overview of the e-mobility demand estimation framework.}
\label{fig:eMobilityDemandFactors}
\end{figure}

Estimations are calculated at the granularity of a London borough. Figure \ref{fig:eMobilityDemandFactors} provides an overview of the framework and its modules (orange frame rectangles on the figure). These are used to determine if the estimated charging needs can be satisfied with the current infrastructure. The output is in the form of a \textit{degree of satisfaction}, \textit{i.e.,} percentage of EVs that can charge using the public charging infrastructure. 

\textbf{Estimation of the average daily energy needs of an EV} To estimate the value of the average daily energy needs of an EV, we propose a simple method. We first determine the average number of kilometers an EV drives on an annual basis. We do so by reporting the total number of vehicles on the road in Greater London in a year to the total number of kilometers driven in the region in a year. We then determine the typical energy consumption of an EV per kilometer by taking into account the influence of a number of variables including the impact of temperature on battery management, the driving behaviour, range anxiety (\textit{i.e.,} frequency of a recharge) and drive train efficiency. We obtain the average daily energy needs of an EV by multiplying the average number of kilometers an EV drives on an annual basis by the typical energy consumption of an EV per kilometer, which normalises at the day level.  

\textbf{Estimation of the overall daily charging needs in Greater London} To estimate the overall charging needs in Greater London on a daily basis, we multiply the charging needs of an EV by the total number of EVs on the road in the region. The latter is obtained by aggregating the number of EVs in every borough.


\textbf{Estimation of the distribution of public charging needs across Greater London's boroughs} To determine the distribution of the needs in terms of public charging across the Greater London area, we proceed by defining two types of charging needs: 1) charging needs that are satisfied at the home location of the EV, and 2) charging needs that are satisfied at the destination location of the EV, \textit{i.e.,} where it travels when it is driven. In both cases, we estimate the distribution of the charging needs at the borough level. For home locations, we assume that the more EVs in a borough, the higher the public charging demand in that borough. The rational is that as the number of EVs increases, the reliance on public charging stations also increases (not all drivers can have access to private chargers). We thus report the distribution of EVs across boroughs to the total number of EVs to estimate the charging needs at home locations. For destination locations, we identify key points of interest in the city such as supermarkets, business hubs, health centers, \textit{etc.} and weight the charging needs in a borough to the number of points of interest in that borough (\textit{i.e.,} borough's attractiveness) such as the larger the number of points of interest, the higher the demand. The distribution of the charging needs in Greater London is estimated by combining the estimation of the needs at home locations and at destination locations. 

To estimate whether charging needs can be satisfied by the public charging infrastructure, we simply compare the estimated charging needs in every borough to the charging load that can be supported by public charging stations. 

\subsection{Discussion}

The proposed estimation methods build upon a number of simplifying assumptions. It is first assumed that it is possible to define a typical EV, and hence calculate a representative EV energy consumption profile. In practice this depends on the make of the vehicle and a number of exogenous factors such as the driving style of the user, the topology of the roads, and the weather conditions. This necessitates to develop not a typical EV but instead a set of typical EV behaviours. In addition, the estimation considers that the overall vehicle's miles in an area are equally split between the number of EVs. In practice journey profiles are needed to be representative of differences in how the vehicle is used. Finally, the framework equally weights the attractiveness of the points of interest in a local area (borough here). In practice this needs to be adapted based on actual popularity of destinations that can be determined by consulting drivers. The framework can easily be extended to accommodate updated assumptions, and hence estimation methods.
\chapter{Added value e-charging services: incentives and collaboration models}
\label{sect:services}

\section{Accommodating e-charging services through incentives}
\label{sect:incentives}


Policymakers have been deploying different policy measures to promote the acceptance of EVs. As a result, the adoption of EVs has been exponentially rising since 2010 with decreasing cost of producing batteries and growing awareness of public grants for ultra-low-emission vehicles among the contributing factors \cite{iea21Report}. An earlier switch by combustion-vehicle drivers (to a less polluting mode) has the value of preventing emissions cumulatively over time. The ability of building an adequate and user-friendly e-Mobility infrastructure is a key part of getting people to switch to more sustainable transportation modes. In that direction, incentives can help with promoting e-Mobility further.

\subsection{How can incentives help?}
\label{sect:how}

Promoting user behavior that improves overall system effectiveness and convenience \cite{dixon2020ease} requires not only monetary incentives and disincentives, but also identifying which factors drivers value, taking advantage of these preferences through “push and pull” measures. Both incentives and disincentives have strengths and weaknesses: disincentives (\textit{e.g.,} taxes, fines) come with the associated expense of enforcement, whereas incentives (\textit{e.g.,} rewards, prizes) imply the cost of rewarding participants per action taken. Thus, whereas incentives become more expensive to administer as they grow, for disincentives the opposite is usually true. 

To investigate the benefits and impacts of such policy scenarios to encourage EV use is crucial before implementing new schemes in real-world. The results of such an analysis can on one hand provide insights on the feasibility of these schemes and on the other hand serve to develop guidance for policymakers on the incentives and disincentives that have the potential to be cost-effective, enabling informed decision-making and saving resources.

e-Mobility users can be encouraged to do something either by making it cheaper, easier or more socially rewarding (other signals may exist), and for each user the intensity of each of these signals required to induce action varies. This means that to determine how sensitive users are to each of these nudges makes it possible to send the right signal (or combination thereof) of the right strength on an individual basis or to groups (as defined according to some external characteristic, \textit{e.g.,} EV ownership or level of energy consumption). Additionally, it is also possible to influence behaviour by using the opposite signals, making anti-social behaviour more costly, more difficult, or more socially unfavourable, as it is entirely possible that the sensitivity of users to positive and negative nudges is not perfectly symmetrical, but rather, that they are more susceptible to one or the other. 

\subsection{Incentives and charging schemes}
\label{sect:impactKEincentives}

We review three forms of charging schemes that different incentives can support, specifically reserved parking for EVs, queuing for charging and charging network operator roaming. An overview of the schemes are given in Table \ref{tab:table_3}. 

\begin{table*}[h]
    \caption{Incentives and charging schemes}
    \centering
    \begin{tabular}{p{2cm}p{4cm}p{4cm}p{4cm}}
        \hline
        \textbf{Scheme name} & Reserved Parking & Queuing for Charging & Operator Roaming \\ \hline
        \textbf{Description} & Reserving on-street parking near charging stations for EV only & Queuing physically or virtually for public charging stations & Allowing inter-charging network access to any service subscriber \\ \hline
\textbf{Mitigation aim} & Anxiety around charging space & Anxiety around charging availability & Anxiety around on-route charging \\ \hline
\textbf{User incentive} & To leave the vehicle near a station and start automatically charging when previous charge terminates. & To plan a journey by making advanced arrangements for charging stations. & To get access to a larger number (ideally all) of charging points on a journey. \\ \hline   
    \end{tabular}
    \label{tab:table_3}
\end{table*}

\textbf{Reserved Parking} The idea behind this scheme is to reduce the anxiety associated with the availability of charging stations by allowing the reservation of parking near charging stations exclusively for EVs (\textit{e.g.,} \cite{latinopoulos2017response}). Such exclusively available parking makes it possible for a driver to leave the car near a station and move it later on for charging, or if and when technically possible, connect to a charging station currently in use via an extension cord that can then supply power when the previous user is done charging. This arrangement also allows equipment able to connect one station to a number of cars for distributed overnight charging. 


\textbf{Queuing for Charging} The rational for this charging scheme is to improve the utilisation of chargers by setting up time-slots for advance booking (\textit{e.g.,} \cite{cao2015reservation}). Not only does this prevent drivers from making trips to chargers without knowing if they are available, it is also a way for recording future demand on a short time basis (\textit{e.g.,} minutes), which is useful for managing the operation of the charging network. Reservations can be made with different levels of economic commitment (i.e. so called hard or soft reservations). The disadvantages for operators to implement such an approach is minimal as reservations can come with an expiry time, and a simple cap on booking cost can prevent users from being priced out.

Another approach is to implement a queuing system, for instance with mobile alerts when the charging station becomes free. The challenge with queuing is that drivers cannot be expected to stay by their vehicles while they wait their turn to charge. For them to perceive queuing as beneficial they might have to know how many people are ahead of them in the queue. Many factors need to be combined to define the cost of implementing a queuing approach and whether this can achieved through software, hardware, and with support from a website or an app. In addition to preventing congestion and energy expenditure of drivers searching for charging, managing queues offers the advantages of allowing better predictions of congestion and electricity grid-loading based on time and place and, of providing a platform for affecting time of charge to make demand as constant as possible. 

The implementation of a queuing approach depends on where the reservations are made (either online or at the charging station), the flexibility of the commitment (hard or soft reservations) and the strength of the price signal, \textit{e.g.,} hard reservations might have to price out the possibility of users exploiting the service. The core idea is that by recording charging instances and requests on a platform, both drivers’ use of infrastructure, and managers’ operation of the infrastructure can be improved, adapting the network to particularly stressed areas.

\textbf{Operator Roaming} The current state of the public charging network coverage at the national and international levels is fragmented, with operators of these networks mostly competing at local and regional levels, and several actors -generally energy providers or petrol stations- offering charging at petrol stations on motorways. This is a limitation for autonomy of EV drivers outside of their home energy networks. As EV uptake grows, EV drivers wishing to travel longer distances are likely to select charging operators providing wider coverage, naturally leading to network consolidation (\textit{e.g.,} through initiatives such as EV Roam in the UK or evRoaming4EU in the EU). In the current context where users remain atomized, an initial approach is a platform where users of different charging networks connect to offer each other access through their subscriptions (in exchange for credits). Another approach is to enable national and international collaboration in order to bring about incentives for network operators of varying scales to coalesce or to enable roaming \cite{mustafa2014roaming}\cite{ferwerda2018advancing}, for instance by allowing members of other networks access to charging for a small surcharge, as done today in the context of mobile communications. This can simplify charging on the users’ side and hence contribute to bolstering EV adoption \cite{noyen2013electric}\cite{dattaMScReport21}. The objective is to implement a system wherein access to as many charging points as possible can be made as simple as possible both for current and prospective EV users.\\

\section{Reversible charging strategies - Vehicle to Grid integration}
\label{sect:smartCharging}

In an effort to tackle climate change, the share of intermittent renewable energies in the energy mix is increasing while the transportation sector is undergoing electrification. The incorporation of millions of EVs on the road is however feared to create extra tension on the electricity network. The increasing share of intermittent renewable electricity sources to replace dispatchable conventional power plants already makes constant balancing between supply and demand of electricity a challenging task, while the lower load factor of renewables and the higher electricity demand call for the reinforcement of the electricity network \cite{IRENA}. The smart integration of electric vehicles is therefore crucial. 

As flexible electricity storage systems, the batteries of a large EV fleet have the potential to provide ancillary services to the electricity network and help incorporate more renewable energy sources. The combination of smart charging and vehicle-to-grid (V2G) schemes can offer a dynamic and environmental solution to meet the growing demand and reduce the need for grid reinforcement \cite{IRENA}. The large scale implementation of these mechanisms raise however significant technical and socio-economic challenges, especially in terms of demonstrating that these approaches work well in real-life scenarios when taking into account grid limitations, battery degradation, market characteristics and individual behaviours.  

\subsection{Smart charging and V2G integration - some definitions}
\label{sect:smartChargingPresentation}

Smart charging refers to the ability of consumers or the grid operator to delay the recharging of electric vehicles away from times of high demand. While most car owners will typically plug in their cars right after getting home from work, smart charging allows the activity to be spread out over the whole night rather than concentrate it at the end of the day, thus, effectively shaving demand peaks \cite{strbac10}. 

V2G operations refer to an upper-level integration of the network, where V2G-enabled electric vehicles can inject electricity into the grid at a period of high demand and/or low generation. Such scheme can help flattening the demand curve, hence reducing the capital costs needed to reinforce the network, and aid with frequency regulation by reducing the need for renewable output curtailment \cite{oldfield21}. From the consumer side, as participants in V2G operations would be paid, the service can incentivise EV ownership and subsequently help further reduce greenhouse gas emissions in transport.

\subsection{Technology and infrastructure requirements}
\label{sect:requirements}
 
V2G operation is based on a bidirectional concept, under which only minor changes to the EV design is needed to include several smart electronic devices. Two main connections are required to enable V2G operations. The first one is the power connection to deliver the energy, from and to the vehicle. The second one is control and logic connections to signal the power demand and the power flow direction \cite{habib2015impact}. On-board reversible battery charger can satisfy this function. 

A prototype of an on-board bidirectional battery charger is presented by Pinto \textit{et al.} in \cite{pinto2013bidirectional}. It uses two power converters that share a DC link: (1) the full-bridge AC-DC bidirectional converter that interfaces the power grid and (2) a reversible DC-DC converter that interfaces the batteries. During the V2G mode of operation, the full-bridge AC-DC bidirectional converter functions as an inverter to convert DC to AC. Synchronization between the converter and the power grid fundamental voltage is required so that the system can work based on the current standards. This is achieved by implementing some logic in a digital controller and determining the reference current that is needed to produce the reference voltage. In order to make the energy flows back from batteries to the grid, the voltage needs to be higher than the power grid voltage in the peak value. Therefore, the DC-DC converter acts as a step-up converter to raise the voltage magnitude. The work provides a proof-of-concept in simulation but in practice, adjustment to the design of the power converter are required for the technology to be seamlessly integrated within an EV and widely accepted to allow a large deployment. 

Most current studies on V2G service build upon the assumption of a large global communications system, which does not exist in today's grid infrastructure. V2G operations extensively depend on the nature of the charging infrastructure. In particular, better information privacy guarantees and scalability can be achieved with a decentralised charging infrastructure but at the cost of higher management complexity of the grid. In contrast, EVs aggregated on centalised public charging stations can ease operations but to the extent of the convenience and privacy of individual vehicle owners. 

Aggregators are likely to become essential actors in that picture as they would allow the exchange of both information and electricity. They could ensure as well that the interests of all stakeholders are met when optimising charging and providing ancillary services.

\subsection{Overview of proposed V2G strategies}
\label{sect:proposedV2GStrategies}

A key challenge for the implementation of V2G is the operation of the power grid at the distribution level. Many studies that have been conducted to probe the feasibility of V2G, do so under specific assumptions and parameter settings. In addition, efforts reported in the literature have been investigating various types of objectives that a V2G strategy can follow. For example, some studies focus on the implications of V2G if implemented country-wide \cite{staudt2018using} by investigating the total demand and total generation capacity limitations. Others focus on optimising for specific grid distribution networks \cite{gao2014integrated} or go beyond typical optimisation objectives such as the minimisation of costs and carbon emissions to discuss the effects of consumer behaviour based on perceptions of V2G technology \cite{wolinetz2018simulating}. 

Table \ref{table:objectiveFunctions} and Table \ref{table:factors} provide an overview of the main objectives and methods proposed in the literature for the implementation of V2G strategies. Today's most comprehensive models incorporate grid, battery, market and population considerations to investigate the potential of these strategies.

\begin{table}[!t]
\caption{Objectives and V2G methods (based on \cite{wang2016quantifying}).}
\label{table:objectiveFunctions}
\vspace{-2mm}
\renewcommand{\arraystretch}{1.2} 
\renewcommand{\tabcolsep}{0.15em}
\small
\centering
\begin{tabular}{|c|c|}
\hline
\textbf{Objective} & \textbf{Method}\\
\hline
Price & Tariffs and operation and consumer costs control \cite{gao2014integrated,wolinetz2018simulating,gree2020cloud}\\
\hline
GHG Emissions & Renewable penetration \cite{staudt2018using}\\
\hline
Energy losses & Distribution network efficiency \cite{staudt2018using}\\
\hline
\multirow{3}{*}{Load flattening} & Valley filling\\
& Peak shaving \cite{gao2014integrated}\\
& Load shifting \cite{staudt2018using}\\
\hline
\multirow{3}{*}{Service quality} & Frequency regulation \cite{gao2014integrated}\\
 & Voltage regulation\\
 & Load factor\\
 & Active power\\
 & Reactive power\\
\hline
\end{tabular}
\end{table}

\begin{table}[!t]
\caption{Main technological, economic and social factors of V2G per category}
\label{table:factors}
\vspace{-2mm}
\renewcommand{\arraystretch}{1.2} 
\renewcommand{\tabcolsep}{0.15em}
\small
\centering
\begin{tabular}{|c|c|}
\hline
\textbf{Category} & \textbf{Factors}\\
\hline
\multirow{2}{*}{Grid} & Current and voltage limits \cite{gao2014integrated}\\
 & Transformer limits\\
\hline
\multirow{4}{*}{Battery} & State of Charge (SOC) \cite{staudt2018using,gao2014integrated,gree2020cloud}\\
 & Desired SOC \cite{gao2014integrated,wolinetz2018simulating,gree2020cloud}\\
& Battery constraints \cite{gree2020cloud}\\
& Battery degradation \cite{gree2020cloud}\\
\hline
\multirow{3}{*}{Market} & Day-ahead prices\\
& Real time prices \cite{gao2014integrated}\\
& Reserve capacity prices \cite{staudt2018using,gao2014integrated}\\
\hline
\multirow{2}{*}{Population} & Homogeneous \cite{staudt2018using}\\
 & Heterogeneous\\
\hline
\multirow{3}{*}{Data} & Collection, storage, analysis \cite{wolinetz2018simulating,gree2020cloud}\\
& Privacy \cite{wolinetz2018simulating}\\
& User interface \cite{gree2020cloud}\\
\hline
\end{tabular}
\end{table}

\subsection{Technical and business limitations}
\label{sect:v2gLimitations}

In spite of positive evidence supporting the technological feasibility of V2G, there are still multiple challenges to overcome in order to reach market maturity.  

First of all, while battery modelling studies suggest that V2G operations could involve higher battery degradation rates, it is not highly distinct from what is expected with uncontrolled charging mechanisms. As such, solutions such as peak shaving, frequency regulation and net load shaving do have a practical feasibility \cite{calvillo2016vehicle}. For load shifting to be profitable, however, the cost of degradation or the cost of the batteries themselves need to be reduced \cite{dubarry2017durability}. The implementation of algorithmic intelligence within smart grids and of Battery Management Systems (BMSes) can enable V2G services to extend the life span of batteries \cite{saldana2019electric} and spread the discharging cycles along the EV population to allow a maximum of 20 discharges per year \cite{calvillo2016vehicle}. In addition, manufacturers are likely to adjust their production to the suitable lithium iron phosphate batteries, which have been proven to be the most adequate for V2G \cite{wang2016quantifying} \cite{crawford2018lifecycle}.

A better understanding of the behaviour of batteries in V2G operations can help design incentives that encourage owners to participate in V2G and thus resolve potential network congestion. A possible strategy could be based on the level of battery degradation, by which the money stream would be redistributed as function of the expected marginal generation cost to redispatch the energy. For instance, using rough estimates, Staudt \textit{et al.} \cite{staudt2018using} evaluated that EV owners would need to be paid at least 4.12 euros per MWh injected into the grid.  

Another challenge of applying V2G in a large scale is data collection and vehicle management. Strbac \textit{et al.} \cite{guille2009conceptual} investigate the implementation of a central management, termed as an “aggregator”, in charge of collecting the vehicle and grid data and balancing the electricity flow via communication with EV consumers and the national grid. Interfaces between the system operators necessitate accurate electricity price prediction that can be obtained with learning-based approaches \cite{chang2019electricity}. This does however require the availability of large volumes of data in real time. To handle this large amount of data, it is suggested that all EV are connected to a common cloud \cite{gree2020cloud}, which enables both the EV owner to maximise profit from V2G connection and the aggregator to manage the charging/discharging schedule. 
\chapter{Connected and automated mobility: security and data protection}
\label{sect:securityPrivacy}

The Internet of Things (IoT) has enabled the development of the Internet of Vehicles (IoV). A connected vehicle utilises a variety of sensors to monitor and map the surrounding environment, to identify speed and distances and to communicate with other vehicles and share information. These processes accumulate substantial amounts of data. With the aim of engineering safer, smarter and more reliable vehicles, the data are processed and shared with the manufacturer to help improve vehicular operation algorithms. Data is also shared with the connected infrastructure that supports road mobility and parking.  

\section{Impact of new security standards for connected and automated cars}
\label{sect:security}

The digitisation of the infrastructure inside and around transports enables new business models in the mobility sector and enhances the utility for the end users whilst incentivising the shift away from fossil fuels. 
Despite the clear benefits to the users, the transport sector and the nation as a whole in terms of  increased service availability and comfort, reduced cost, job creation and sustainability, the challenges that need to be tackled to protect this critical digital infrastructure against cyber-threats can be overlooked. 
As we build those new services of the future, safety and security must be considered by design rather than included as an afterthought. This is not an easy task to achieve given the complexity of the systems involved and the heterogeneity of the devices that need to be interconnected to enable new valuable services to customers.  

\subsection{Challenges for future vehicle connectivity and data security}

Whilst most modern day vehicles support internal networks of electronic control units with control over vehicular function, the level of connectivity associated with the shift of connected and automated vehicles to the mainstream is unprecedented~\cite{security}. The rise of Electric Vehicle Cloud and Edge Computing (EVCE) provides an attractive opportunity for companies to reduce pressure on central cloud computing power and enables local aggregation of vehicle's data to ensure anonymity at the cloud processing level \cite{Liu2018}. However, by expanding the computing power to the edge, to the connected vehicles, it is increasing the system surface area vulnerable to attack. All of the network participants must be protected against attack, as well as the cloud system, or large amounts of data could be leaked to one network vehicle being compromised~\cite{edge}. 

Greater levels of vehicle's connection to home smart systems also raise security challenges. Any smart device on the home network can fall to a malicious attack and consequentially affect all other devices, including connected vehicles \cite{Martinez}. Malware could then be transmitted to other networks, such as public charging networks, impacting the confidentiality and integrity of information exchanged.

A prominent feature of EVCE computing models is the increased vehicle-to-vehicle (V2V) communications which enables information, such as traffic and road environment conditions, to be shared. This information is reliant on vehicle spatio-temporal data for processing, defined as sensitive data. As stated previously, with a greater number of network nodes, the vulnerability to attack does increase if one node/vehicle is compromised. Protecting this information whilst also ensuring traceability of interactions to reduce risk of unnoticed malicious manipulation presents a security trade off problem. Research into the use of blockchain as a remedy is ongoing~\cite{Liu2018}. 

Compromise of the V2V system can lead to jamming or denial-of-service attacks where essential information exchanges are rendered unable to occur, impacting data availability. Exchanges range from sensor information and traffic updates to the electronic control units within the vehicle and ability to report issues to the central cloud server. False data injections into the V2V system are also a concern as they may mislead drivers about battery levels or temperature which can be transmitted back to the central system and impact its predictive and decision recommendation abilities~\cite{jam}. Deployment of 5G and the introduction of new encryption methods provide opportunities to ensure better security of communications between the vehicle, sensors, and the network which are achieved via cellular, satellite and WiFi services~\cite{cell,5g}.

Finally, the increase of connected vehicles in the mainstream could see a rise in phishing and malicious software downloads. Software and firmware updates should come from trusted sources to avoid malicious software being downloaded, infecting the network and resulting in system data leaks~\cite{security}. This is an example of targeting the human factor in the connected and automated mobility ecosystem, and car's companies need to be careful to warn customers of such risks and produce plans to mitigate the impact of one vehicle being affected to the system.

\subsection{Why are standards and regulations needed ?}
\label{sect:whyStandards}

\subsubsection{Software vulnerabilities}

It is estimated that it takes 100 million lines of code to get a modern vehicle rolling \cite{spectrumCharette09}. Commercial software typically has 15 to 50 bugs per $1,000$ lines of code \cite{sandu2018new}. A rough estimation using these numbers point to the existence of hundreds of thousands of potential vulnerabilities, some benign others more critical, waiting to be found and exploited by a malicious actor. Increased connectivity in the transport infrastructure facilitates access to and exploitation of those vulnerabilities. 
The digitisation of the transport ecosystem increases the likelihood of occurrence of a cyber incident that could disrupt major services with potentially tragic consequences in terms of loss of life or environmental damage along with serious economic losses. Ironically, these are the very outcomes that the digitisation of transport infrastructure is set to prevent. 
A report by Upstream \cite{upstreamReport}, a leading automotive cybersecurity firm, shows that over the period 2010-2020, 49.3\% of attacks that the firm analysed were due to black hat hackers versus 45.8\% from white hat hackers\footnote{The term Black hat hackers denotes hackers with malicious intentions whereas White hat hackers are usually security researchers rewarded for reporting security vulnerabilities to the company issuing a given product.}. In 2020, 54.6\% of incidents were attacks by black-hat hackers, which hints at a steady increase in the number of attacks from malicious actors. More concerning is that almost 80\% of the attacks analysed between 2010 and 2020 were performed remotely through WiFi, Bluetooth 3/4/5G, \textit{etc.} Although less prominent, physical attacks are also a major concern that requires attention, especially as we move towards car sharing type of services. The next customer to use a vehicle should be able to assess with certainty that the vehicle was not compromised in any way by the previous customer. 

\subsubsection{A complex ecosystem}

Securing connected vehicles and the transport infrastructure around it is made even more challenging due to the fragmented and complex supply ecosystem that is relied upon by this industry. Vulnerabilities in the products delivered by Tier-1/Tier 2 suppliers to OEMs (Original Equipment Manufacturer) can be exploited to gain control of a vehicle. For example in May 2020 a vulnerability in ARMv7 implementation of GNU C library allowed hackers to perform remote code execution on a connected vehicle \cite{may20Incident}. These types of vulnerabilities can affect multiple vendors as the same systems are often used by different OEMs in their respective lines of products. 
In addition to the vulnerabilities present in the software of the vehicle itself it is also crucial to pay attention to the infrastructure around the vehicle, be it the fleet management servers or the increasingly smart traffic management systems that are moving towards higher integration with the vehicles systems by providing real time information to the vehicles’ decision making units. A scenario whereby a compromised road sensor is used to disrupt traffic is not in the realm of science fiction. Such disruptions can also be experienced with human drivers faced with a dysfunctional traffic light. It is easy to extrapolate what would happen if automated systems were made to rely on fraudulent data. In \cite{10.5555/2671293.2671300} Ghena \textit{et al.} analyse a networked traffic system deployed in the US. They successfully identify and exploit vulnerabilities that they further use to gain control of the traffic infrastructure. In the future of Intelligent Transport Systems (ITS) this type of vulnerabilities can cause massive disruptions. These examples constitute evidence that this is a critical national security matter. 

\subsubsection{Privacy concerns}

Coupled to the security concerns previously mentioned are the data privacy concerns that come with any form of digitised platform. Questions remain unanswered regarding the ownership, custody and exploitation of the data produced by the various devices connected to the transport network and their interaction with members of the public. Important questions arise from the availability and use of data related to passengers travel habits. Hahn \textit{et al.} define seven types of privacy of which \textit{privacy of behaviour} and \textit{privacy of location and space} are clearly threatened by advances in ITS \cite{hahn2019security}. Although data privacy regulations such as the General Data Protection Regulation (GDPR) framework (see Section \ref{sect:gdpr}) are now becoming widely implemented, it is still a challenge to understand how these can effectively be applied in a context where public data is not necessarily in the hands of a centralised entity but rather accessible to the masses. 
 
\subsection{Steps taken to mitigate cyber risks}
\label{sect:mitigation}

Given the safety risk posed by cyber security issues in the transport industry, regulators have come together to provide specific legal requirements and standards that will hopefully drive the entire industry towards the cyber security best practices. 

The UNECE World Forum for Harmonisation of Vehicle Regulations (WP.29) \cite{wp29} has produced new regulations focusing on cyber security and Over-The-Air (OTA) software updates in the context of road vehicles.  This regulation requires OEMs to build certified Cyber Security Management Systems (CSMS) and provides guidance as to the implementation of such a system. For example a CSMS must be risk-based, \textit{i.e.,} it needs to take into account the specific risks of the systems built by the OEM. This is to avoid OEMs to adopt a checkbox approach to security and purchase blanket security solutions from vendors irrespective of their systems architecture. This CSMS must also be audited by a third party for compliance with the regulation. This regulation is successful if it incentivises OEMs to consider cyber security in the design phase of their product lifecycle. Additionally, OEMs are required to implement detection and prevention mechanisms similar to Intrusion Detection/Prevention Systems (IDS/IPS) common in IT infrastructure. These IDS/IPSes need to be specific to the context of road vehicles (\textit{e.g.,} IDS system developed by Abba-IoT \cite{abbaiot}) . Emphasis is also put on the capability to perform forensics analysis post incidents. This is to inform a better design of security systems and a continuous progress towards increased security.

In addition to regulations, new standards have been released to help OEMs comply with the new regulations such as ISO/SAE 21 434 on Cyber security for Road Vehicles. This standard serves as a reference implementation of the UNECE regulation. It is however encouraged that OEMs do not view it as a checklist for their security analysis but as a methodology to design their CSMS based on their own specific risks.

\subsection{Indirect impact of the regulation and standards}
\label{sect:standardsImpact}

The formulation of regulations and standards for cyber security in the automotive sector is an important development because it compels OEMs to consider cyber security from the design phase of their product and to adopt the best practices to ensure that safe products are delivered to the customer. Although a limitation of this regulation is that it does not apply directly to the OEMs suppliers, the regulatory pressure felt by the OEMs can be passed down to the entire supply chain and incentivise them to adopt higher standards and practices when it comes to cyber security. 
Another indirect consequence of the UNECE regulation is in stimulating the industry to develop capabilities that are key in dealing with cyber vulnerabilities. This includes the building of innovative solutions, for example for IDS/IPS, that are not only a copy and paste of IT versions but that takes into consideration the unique aspect of cyber-physical systems. It also involves training specialists that understand the idiosyncrasies of transport technology, the risks, environment and interconnection with the IT world. This might prove to be a challenge in the short term as there is already an important skill shortage of technical cyber security experts but only by making cyber security a priority can this issue be resolved in the long term.

\section{Implications of legal frameworks on connected and automated mobility}
\label{sect:privacy}

A connected vehicle is able to process up to 19 Tb of data per hour \cite{Tuxera}. This leads to a tremendous amount of collected information. In Europe, how this data can be handled, stored and shared is constrained by the GDPR \cite{GDPR1}, a framework that aims to protect the data of individuals. For instance, Tesla, which has vehicles operating a series of highly complex algorithms to gather data, have recently experienced allegations concerning how the data are handled \cite{Elephant}. Regulation for Tesla's vehicles are set under specific US state laws and do not necessarily comply with the European GDPR. Hence, when these vehicles are on European ground, there are risks of regulatory non-compliance. 
\subsection{What type of data is being collected?}

A connected vehicle, as a part of the Internet of Vehicles, is a highly computerised machine; a vehicle that interconnects with the surrounding environment through a complex, integrated network \cite{Chen}. The vehicle is reliant upon interpreting data to be able to connect and communicate with other vehicles, sensors, charging points, and road infrastructure. The result of this is that a substantial amount of data are being continuously gathered, ranging from 1.4 Tb up to 19 Tb per hour \cite{Tuxera}. Data are gathered specifically through camera and radar technology. Central to these collection processes is the constant access ability to big data. Data from sensors are directed to a controlling unit within the vehicle that processes and handles information required for maintaining vehicular operations \cite{Huang}. The data can be shared to other connected vehicles or sensors through the concept of V2X, acting essentially as the link the vehicle has to its surrounding environment at any given time \cite{Fraiji}. This process is typically done through an intra-vehicular wireless sensor network \cite{Lin}, with the main purpose of identifying potential dangerous situations \cite{Chen}, and to improve the autonomous algorithm. 

A substantial amount of data are continuously gathered via the multiple vehicular sensors with autonomous driving features. This data range is highlighted in Table \ref{tab:DataHandling} and includes how the data from each source is handled via cellular, WiFi or satellite. 

\begin{table}[ht]
\centering
\caption{Connected vehicle's data source and handling}
\includegraphics[width=\textwidth]{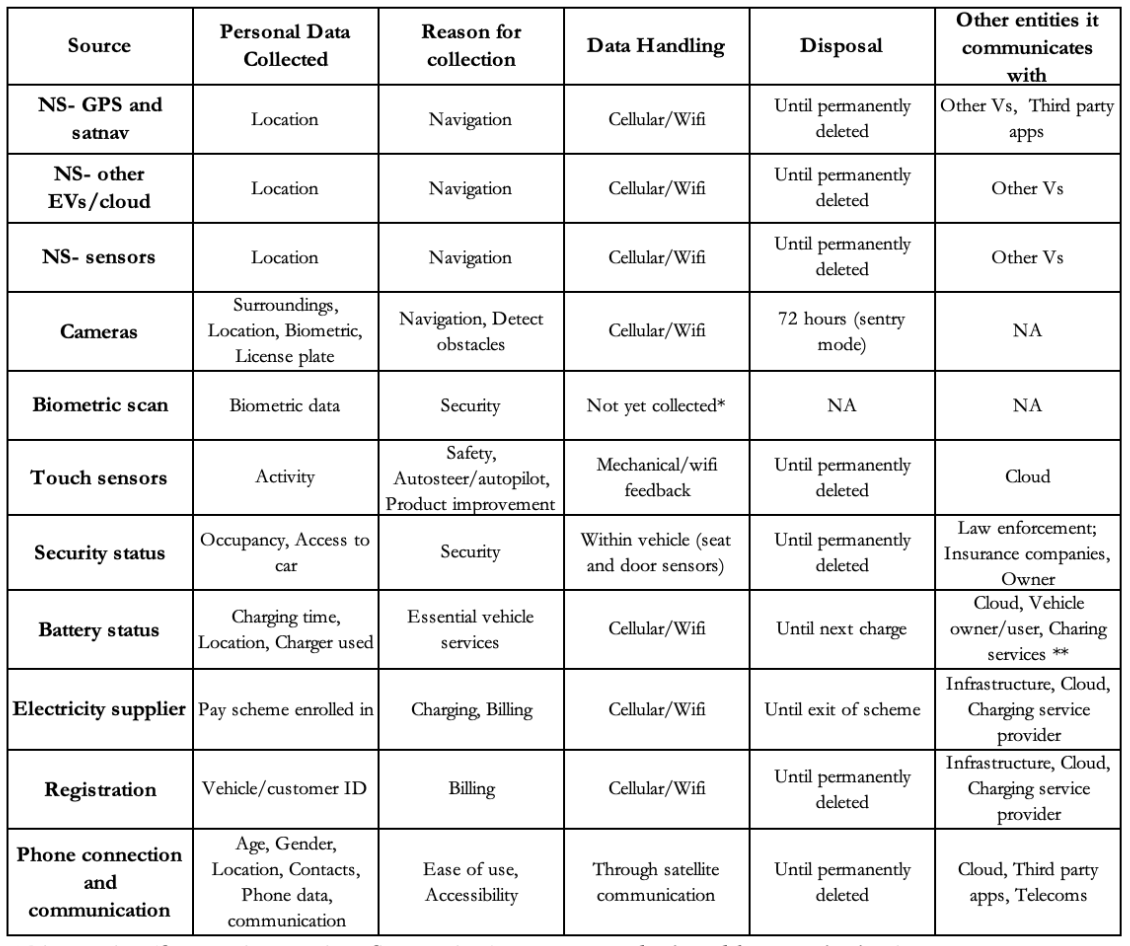}
\label{tab:DataHandling}
\end{table}

\subsection{The General Data Protection Regulation (GDPR) framework}
\label{sect:gdpr}

\subsubsection{A short history of the GDPR} 
"Everyone has the right to respect for his private and family life, his home and his correspondence." - European Convention on Human Rights. In Europe, the right to privacy was initially covered in the 1950 European Convention on Human Rights \cite{ECHR}. The emergence of the Internet led to the establishment of the European Data Protection Directive in 1995 \cite{GDPREU} that primarily focused on setting minimum data privacy and security standards to be implemented by each member state. The evolution of technology and the transition to a digital era highlighted the importance of more stringent laws to protect a vulnerable population from data collecting entities who, in numerous instances \cite{NYT}, unfairly capitalise on personal data for commercial and/or political purposes. 

In light of this, the European Union (EU) created the General Data Protection Regulation (GDPR) in 2016, implemented in May 2018, to replace an obsolete Data Protection Directive \cite{hgdpr}. The new regulation provided "A comprehensive approach on personal data protection" fit for the time; it imposes obligations on all EU operating organisations to legally collect, process, store, and protect all EU residents' personal data \cite{GDPR1}. The regulation defined personal data beyond an individual's name, contact details and location to include IP address, race, sexual orientation, political opinions, religious beliefs, biometric data, etc \cite{GDPR1}; increasing people's privacy rights. It is defined as the most rigid privacy, and security law in the world \cite{GDPREU}. 

\subsubsection{Context of the GDPR} 
The GDPR is composed of 7 guiding principles including: lawfulness, fairness and transparency; purpose limitation; data minimisation; accuracy; storage limitation; integrity and confidentiality (security); and accountability \cite{GDPREU}. 

If any of the above mentioned principles are infringed, the organisation receives a fine of 20 million euros or 4\% total annual turnover (whichever is the higher amount).  In enforcing the GDPR, companies such as Google and British Airways have been fined 50 million euros for an insufficient legal basis for data processing and 20 million euros for insufficient organisational and technical measures to ensure information security respectively \cite{CNIL,ICOBA}. 
 
\subsubsection{GDPR, EVs and connected cars}
IoT is enabling the transport sector to innovate and increase in complexity by shifting from a hardware to a software focus. As a consequence, the legal frameworks necessary to ensure a fair, equitable and inclusive transition to Intelligent Transport Systems \cite{Dictionary} are evolving to include data protection and cyber security \cite{Costantini2020}. The primary objective of transport regulations is road safety, which requires "accessibility, exchange and re-use" of specific traffic and road data as per the Delegated Regulation (EU) 2015/962 \cite{Delegated}. The incorporation of GDPR in national legislation may present challenges for transport policy innovation, an issue already highlighted by the legal battles and fines encountered by novel mobility services such as Uber. Provisions for the management and protection of Autonomous and Connected Vehicles data (AVs and CVs) are still at the initial stages across EU countries \cite{Costantini2020} though national and transnational progress has been made through the use of regulatory sandboxes ("Living Labs") \cite{Livinglabs}.

Tesla, founded in 2003, is a pioneer in the commercialisation of electric vehicles \cite{AboutTesla}, with its Model 3 being the highest selling plug-in car in the UK in 2020 \cite{ZapMap}. However, data handling in Tesla vehicles such as the Model 3 has been found to breach Art. 5(1)(b), Art. 13 and  Art. 14 of the GDPR due to the lack of specification in the purposes of the data processing \cite{Elephant} as well as transfer of user data to locations outside the EU. The continuous video and ultrasonic monitoring that enable Sentry Mode \cite{TeslaSupport} and facilitate driving, present significant concerns for data protection and would need to be disabled on EU member state roads to comply with GDPR.

\subsection{Data privacy characteristics}

Most of the data collected through the sensors of a connected car lie within the GDPR definition of personal data \cite{GDPR1}. The GDPR obligates vendors to obtain the vehicle owner's consent before collecting, processing and storing any data because of its potential sensitivity. \textit{Sensitive data}, which is the majority of personal data, is defined as classified data that must be protected and remain inaccessible to outside parties unless the owner of the data grants permission \cite{CIP}. In the case of Tesla's cars, for instance, these include financial information, racial or ethnic origin, health-related data, religious beliefs and political opinions, which are directly collected or could potentially be inferred from the collected data. While easily amendable, it is worth noting that Tesla's terms and conditions have been spotlighted as an issue as it does not explicitly state the reason for data collection \cite{Elephant}.

While trying to improve their self-driving options, car manufacturers face an additional challenge as the built-in cameras used to collect data for training their neural networks also record non-consenting individuals' data (biometrics and vehicle details). Tesla vehicles for instance have been regarded as "surveillance machines" in Germany, where they have received continuous backlash \cite{Elephant}.

\begin{table}[ht]
\centering
\caption{Risk of data handling from source}
\includegraphics[width=200pt]{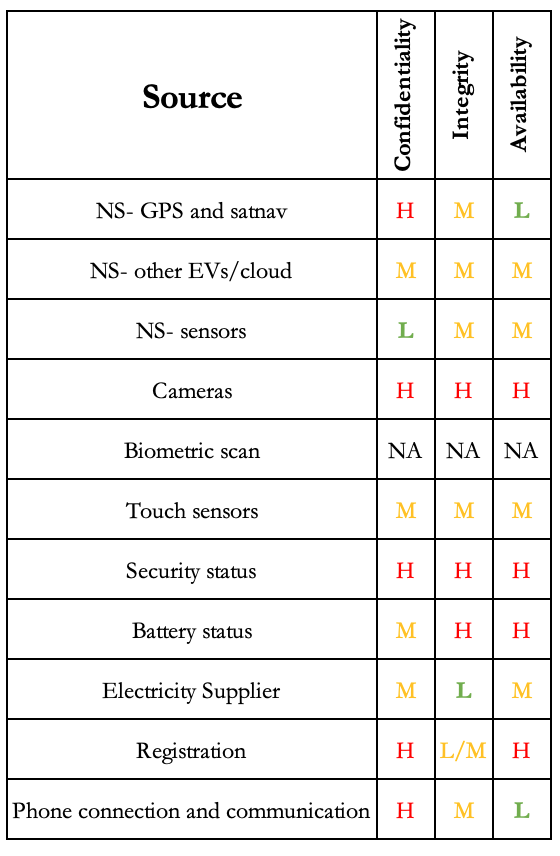}
\label{tab:RiskOfSource}
\end{table}

While all personal data can be defined as sensitive, the degree of sensitivity differs as not all data is created equal, and the legal ramifications as a direct result of a breach differ between data types. Therefore, it is imperative for vehicle's vendors to assess the varying degree of sensitivity of the data collected to be able to comply with the GDPR's data protection obligations. We focus on Confidentiality, Integrity and Availability (CIA) properties of a data source to define the level of data sensitivity in the case of connected and automated cars. 

\begin{itemize}
	\item \textbf{Confidentiality} is about protecting data against unauthorised access, disclosure, or theft and is directly linked to information privacy \cite{Jouvray}. Information with low confidentiality concerns may be considered less threatening if exposed beyond its intended audience, while one with high confidentiality concerns is considered private and must be protected against malicious intents \cite{DEL}.
	\item \textbf{Integrity} assesses protection of data exchange and data storage against improper modifications or destruction \cite{Jouvray}.
	\item \textbf{Availability} \cite{Jouvray} is about ensuring a timely, continuous and reliable exchange and use of data.
\end{itemize}

In Table \ref{tab:RiskOfSource} we qualitatively assess the potential impact on the operations, assets, and reputation of a vendor if there was to be a confidentiality, integrity or availability violation. The sensitivity of the data is directly related to the potential impact level, with low impact indicating a low sensitivity. As highlighted in the table, in the context of the GDPR, confidentiality, integrity, and availability are crucial properties that vendors need to comply with when handling data. 

\subsection{Discussion}

Connected and automated vehicles often have an interactive interface that presents conditions requiring consent for vehicle operation and safety. The data that the vehicle collects can not be processed without the consent of the driver. However, obtaining consent via this interface can be influenced by software design \cite{Pattinson}. A driver might agree to terms and conditions without understanding the legal implications which follow, or be presented with an unnecessarily long policy which most drivers will not read in full. Consequently, the driver may not comprehend what data is collected.

In the case of Tesla, a video surveillance protection system called Sentry Mode is used to detect movement outside the vehicle. This is incompatible with the current GDPR regulation as the lightly triggered cameras record passers-by and stores the footage for a while even without a determined threat. This footage captures sensitive personal data such as people’s faces and license plate numbers without consent, breaching data privacy. This presents an issue where the automatic camera recording is necessary for Sentry Mode’s protection system. It is however heavily conflicting with the GDPR and non-negotiable because it is at the cost of the data subjects that have not given consent.  


It is also the responsibility of the company to comply with the information requirements of Article 12 of the GDPR \cite{GDPR1} which include an indication of the purpose for data collected, those responsible for the handling and processing of the data and the rights of the data subject/user. Tesla is for instance accused of vaguely and unsatisfactorily addressing these formalities. Through technical and legal measures such as the above-mentioned in the sentry mode instance and a Data Protection Impact Assessment may provide good solutions for GDPR compliance for vehicle's manufacturer. Another potential problem is whether users are able to request for their data not to be used for instance to improve the performance of the company's algorithms, \textit{e.g.,} using data on drivers’ hand placements to improve data sets and user experience.

\section{Concluding remarks}
\label{sect:securityConclusion}

Connected and autonomous vehicles are only the logical next step in the evolution of transport, it comes with a vision of a future where transports are more sustainable, more efficient, more comfortable and safer. To realise this vision it has become urgent to overcome the hurdle of cyber security so that this future does not turn into a dystopia where the entire transport infrastructure of a country is grounded to a halt or lives are put at risk by malicious actors. The new standards and regulations developed specifically around cyber security in transports are leading the way to a brighter future by changing cyber security from being an afterthought to becoming part of the design process of a vehicle.  Now is the right time to act before the infrastructure is built to incorporate security by design and develop the tools and competences necessary to produce and maintain secure systems. It will be important to see in the coming years how the automotive sector has responded to these new rules and hopefully gone above and beyond the basic requirements.

\bibliographystyle{unsrt} 
\bibliography{Report} 

\begin{thebibliography}{10}

\bibitem{obstfeldReport}
J.~Obstfeld.
\newblock Global engineering: Enabling a connected transport future.
\newblock Cisco Blogs, 2019.
\newblock https://blogs.cisco.com/sp/connectedcar-thedrivenhour-wp - Accessed
  01-03-2022.

\bibitem{ferrara2019multimodal}
Marina Ferrara, Carlo Liberto, Marialisa Nigro, Martina Trojani, and Gaetano
  Valenti.
\newblock Multimodal choice model for e-mobility scenarios.
\newblock {\em Transportation Research Procedia}, 37:409--416, 2019.

\bibitem{yang2018user}
Woosuk Yang.
\newblock A user-choice model for locating congested fast charging stations.
\newblock {\em Transportation Research Part E: Logistics and Transportation
  Review}, 110:189--213, 2018.

\bibitem{straka2020predicting}
Milan Straka, Pasquale De~Falco, Gabriella Ferruzzi, Daniela Proto, Gijs Van
  Der~Poel, Shahab Khormali, and Lubo{\v{s}} Buzna.
\newblock Predicting popularity of electric vehicle charging infrastructure in
  urban context.
\newblock {\em IEEE Access}, 8:11315--11327, 2020.

\bibitem{birkel2020challenges}
Hendrik Birkel, Matthias Kopyto, and Corinna Lutz.
\newblock Challenges of applying predictive analytics in transport logistics.
\newblock In {\em Proceedings of the 2020 on Computers and People Research
  Conference}, pages 144--151, 2020.

\bibitem{pluntkeDissertation14}
C.~Pluntke.
\newblock {\em {Modelling users in networks with path choice: four studies in
  telecommunications and transit}}.
\newblock PhD thesis, University College London (UCL), 2014.

\bibitem{lam2014electric}
Albert~YS Lam, Yiu-Wing Leung, and Xiaowen Chu.
\newblock Electric vehicle charging station placement: Formulation, complexity,
  and solutions.
\newblock {\em IEEE Transactions on Smart Grid}, 5(6):2846--2856, 2014.

\bibitem{EvolveProject}
Evolve: Electric vehicle fleet optimisation for lowering vehicle emissions.
\newblock https://www.imperial.ac.uk/evolve-project/ - Accessed 01-03-2022.

\bibitem{LondonCouncilsStat}
London councils - who runs london.
\newblock
  https://www.londoncouncils.gov.uk/who-runs-london/essential-guide-london-local-government
  - Accessed 01-03-2022.

\bibitem{ULEVsStat}
Licensed ultra low emission vehicles by local authority: United kingdom,
  report: Veh0132, table veh0132a.
\newblock
  https://www.gov.uk/government/statistical-data-sets/all-vehicles-veh01\#ultra-low-emissions-vehicles
  - Accessed 01-03-2022.

\bibitem{DfTLicensingStat}
Department for transport, vehicle licensing statistics: Annual 2019.
\newblock
  https://www.gov.uk/government/statistics/vehicle-licensing-statistics-2019 -
  Accessed 01-03-2022.

\bibitem{zapmapStat}
Zapmap ev charging stats.
\newblock https://www.zap-map.com/statistics/ - Accessed 01-03-2022.

\bibitem{tfl1120Report}
F.~Oldfield, K.~Kumpavat, R.~Corbett, A.~Price, M.~Aunedi, G.~Strbac,
  C.~O'Malley, D.~Gardner, D.~Pfeiffer, and J.T. Kamphus.
\newblock London electric vehicle infrastructure delivery plan: One year on
  november 2020.
\newblock Technical Report supported by The Mayor of London's Electric Vehicle
  Infrastructure Taskforce, 2020.
\newblock
  https://tfl.gov.uk/ruc-cdn/static/cms/documents/london-electric-vehicle-infrastructure-delivery-plan-one-year-on-november-2020.pdf
  - Accessed 01-03-2022.

\bibitem{zapmapNetw}
Zapmap public charging networks.
\newblock https://www.zap-map.com/charge-points/public-charging-point-networks/
  - Accessed 01-03-2022.

\bibitem{iea21Report}
International~Energy Agency.
\newblock Policies to promote electric vehicle deployment.
\newblock Global EV Outlook 2021, 2021.
\newblock
  https://www.iea.org/reports/global-ev-outlook-2021/policies-to-promote-electric-vehicle-deployment
  - Accessed 01-03-2022.

\bibitem{dixon2020ease}
James Dixon, Peter~Bach Andersen, Keith Bell, and Chresten Tr{\ae}holt.
\newblock On the ease of being green: An investigation of the inconvenience of
  electric vehicle charging.
\newblock {\em Applied Energy}, 258:114090, 2020.

\bibitem{latinopoulos2017response}
Charilaos Latinopoulos, Aruna Sivakumar, and JW~Polak.
\newblock Response of electric vehicle drivers to dynamic pricing of parking
  and charging services: Risky choice in early reservations.
\newblock {\em Transportation Research Part C: Emerging Technologies},
  80:175--189, 2017.

\bibitem{cao2015reservation}
Yue Cao, Ning Wang, Young~Jin Kim, and Chang Ge.
\newblock A reservation based charging management for on-the-move ev under
  mobility uncertainty.
\newblock In {\em 2015 IEEE Online Conference on Green Communications
  (OnlineGreenComm)}, pages 11--16. IEEE, 2015.

\bibitem{mustafa2014roaming}
Mustafa~A Mustafa, Ning Zhang, Georgios Kalogridis, and Zhong Fan.
\newblock Roaming electric vehicle charging and billing: An anonymous
  multi-user protocol.
\newblock In {\em 2014 IEEE International Conference on Smart Grid
  Communications (SmartGridComm)}, pages 939--945. IEEE, 2014.

\bibitem{ferwerda2018advancing}
Roland Ferwerda, Michel Bayings, Mart Van~der Kam, and Rudi Bekkers.
\newblock Advancing e-roaming in europe: Towards a single “language” for
  the european charging infrastructure.
\newblock {\em World Electric Vehicle Journal}, 9(4):50, 2018.

\bibitem{noyen2013electric}
Kay Noyen, Matthias Baumann, and Florian Michahelles.
\newblock Electric mobility roaming for extending range limitations.
\newblock In {\em ICMB}, page~13, 2013.

\bibitem{dattaMScReport21}
A.~Datta.
\newblock {Quantifying the Incremental Economic and Consumer Value Benefits of
  E-Roaming Within the UK's Future Public EV Charging Ecosystem}.
\newblock MSc Thesis, Imperial College London, 2021.

\bibitem{IRENA}
IRENA.
\newblock Renewable energy integration in power grids technology brief, 2015.
\newblock
  https://www.irena.org/publications/2015/Apr/Renewable-energy-integration-in-power-grids
  - Accessed 01-03-2022.

\bibitem{strbac10}
G.~Strbac, C.~Gan, V.~Stanojević, P.~Djapic, P.~Mancarella, A.~Hawkes,
  D.~Pudjianto, S.~Vine, J.~Polak, D.~Openshaw, S.~Burns, P.~West, D.~Brogden,
  A.~Creighton, and A.~Claxton.
\newblock Benefits of advanced smart metering for demand response based control
  of distribution networks, 2010.

\bibitem{oldfield21}
F.~Oldfield, K.~Kumpavat, R.~Corbett, A.~Price, M.~Aunedi, G.~Strbac,
  C.~O'Malley, D.~Gardner, D.~Pfeiffer, and J.T. Kamphus.
\newblock The drive towards a low-carbon grid: Unlocking the value of
  vehicle-to- grid fleets in great britain, 2021.

\bibitem{habib2015impact}
Salman Habib, Muhammad Kamran, and Umar Rashid.
\newblock Impact analysis of vehicle-to-grid technology and charging strategies
  of electric vehicles on distribution networks--a review.
\newblock {\em Journal of Power Sources}, 277:205--214, 2015.

\bibitem{pinto2013bidirectional}
JG~Pinto, V{\'\i}tor Monteiro, Henrique Gon{\c{c}}alves, Bruno Exposto, Delfim
  Pedrosa, Carlos Couto, and Jo{\~a}o~L Afonso.
\newblock Bidirectional battery charger with grid-to-vehicle, vehicle-to-grid
  and vehicle-to-home technologies.
\newblock In {\em IECON 2013-39th Annual Conference of the IEEE Industrial
  Electronics Society}, pages 5934--5939. IEEE, 2013.

\bibitem{staudt2018using}
Philipp Staudt, Marc Schmidt, Johannes G{\"a}rttner, and Christof Weinhardt.
\newblock Using vehicle-to-grid concepts to balance redispatch needs: A case
  study in germany.
\newblock In {\em Proceedings of the Ninth International Conference on Future
  Energy Systems}, pages 80--84, 2018.

\bibitem{gao2014integrated}
Shuang Gao, KT~Chau, Chunhua Liu, Diyun Wu, and Ching~Chuen Chan.
\newblock Integrated energy management of plug-in electric vehicles in power
  grid with renewables.
\newblock {\em IEEE Transactions on Vehicular Technology}, 63(7):3019--3027,
  2014.

\bibitem{wolinetz2018simulating}
Michael Wolinetz, Jonn Axsen, Jotham Peters, and Curran Crawford.
\newblock Simulating the value of electric-vehicle--grid integration using a
  behaviourally realistic model.
\newblock {\em Nature Energy}, 3(2):132--139, 2018.

\bibitem{wang2016quantifying}
Dai Wang, Jonathan Coignard, Teng Zeng, Cong Zhang, and Samveg Saxena.
\newblock Quantifying electric vehicle battery degradation from driving vs.
  vehicle-to-grid services.
\newblock {\em Journal of Power Sources}, 332:193--203, 2016.

\bibitem{gree2020cloud}
Florent Gr{\'e}e, Vitaliia Laznikova, Bill Kim, Guillermo Garcia, Tom Kigezi,
  and Bo~Gao.
\newblock Cloud-based big data platform for vehicle-to-grid (v2g).
\newblock {\em World Electric Vehicle Journal}, 11(2):30, 2020.

\bibitem{calvillo2016vehicle}
Christian~F Calvillo, Karolina Czechowski, Lennart S{\"o}der, Alvaro
  Sanchez-Miralles, and Jos{\'e} Villar.
\newblock Vehicle-to-grid profitability considering ev battery degradation.
\newblock In {\em 2016 IEEE PES Asia-Pacific Power and Energy Engineering
  Conference (APPEEC)}, pages 310--314. IEEE, 2016.

\bibitem{dubarry2017durability}
Matthieu Dubarry, Arnaud Devie, and Katherine McKenzie.
\newblock Durability and reliability of electric vehicle batteries under
  electric utility grid operations: Bidirectional charging impact analysis.
\newblock {\em Journal of Power Sources}, 358:39--49, 2017.

\bibitem{saldana2019electric}
Gaizka Salda{\~n}a, Jose~Ignacio San~Martin, Inmaculada Zamora,
  Francisco~Javier Asensio, and Oier O{\~n}ederra.
\newblock Electric vehicle into the grid: Charging methodologies aimed at
  providing ancillary services considering battery degradation.
\newblock {\em Energies}, 12(12):2443, 2019.

\bibitem{crawford2018lifecycle}
Alasdair~J Crawford, Qian Huang, Michael~CW Kintner-Meyer, Ji-Guang Zhang,
  David~M Reed, Vincent~L Sprenkle, Vilayanur~V Viswanathan, and Daiwon Choi.
\newblock Lifecycle comparison of selected li-ion battery chemistries under
  grid and electric vehicle duty cycle combinations.
\newblock {\em Journal of Power Sources}, 380:185--193, 2018.

\bibitem{guille2009conceptual}
Christophe Guille and George Gross.
\newblock A conceptual framework for the vehicle-to-grid (v2g) implementation.
\newblock {\em Energy policy}, 37(11):4379--4390, 2009.

\bibitem{chang2019electricity}
Zihan Chang, Yang Zhang, and Wenbo Chen.
\newblock Electricity price prediction based on hybrid model of adam optimized
  lstm neural network and wavelet transform.
\newblock {\em Energy}, 187:115804, 2019.

\bibitem{security}
H.~Chaudhry and T.~Bohn.
\newblock Security concerns of a plug-in vehicle.
\newblock In {\em 2012 IEEE PES Innovative Smart Grid Technologies (ISGT)},
  pages 1--6. IEEE, 2012.

\bibitem{Liu2018}
H.~Liu, Y.~Zhang, and T.~Yang.
\newblock Blockchain-enabled security in electric vehicles cloud and edge
  computing.
\newblock {\em IEEE Network}, 32(3):78--83, 2018.

\bibitem{edge}
Rodrigo Roman, Javier Lopez, and Masahiro Mambo.
\newblock Mobile edge computing, fog et al.: A survey and analysis of security
  threats and challenges.
\newblock {\em Future Generation Computer Systems}, 78:680--698, 2018.

\bibitem{Martinez}
et~al. Martinez, Jaiber.
\newblock Smart grid challenges through the lens of the european genereal data
  protection regulation.
\newblock {\em International Conference on Information Systems Development},
  2019.

\bibitem{jam}
Yosra Fraiji, Lamia~Ben Azzouz, Wassim Trojet, and Leila~Azouz Saidane.
\newblock Cyber security issues of internet of electric vehicles.
\newblock In {\em 2018 IEEE Wireless Communications and Networking Conference
  (WCNC)}, pages 1--6. IEEE, 2018.

\bibitem{cell}
Ning Lu, Nan Cheng, Ning Zhang, Xuemin Shen, and Jon~W Mark.
\newblock Connected vehicles: Solutions and challenges.
\newblock {\em IEEE internet of things journal}, 1(4):289--299, 2014.

\bibitem{5g}
Roger~Piqueras Jover.
\newblock The current state of affairs in 5g security and the main remaining
  security challenges.
\newblock {\em CoRR}, abs/1904.08394, 2019.

\bibitem{spectrumCharette09}
R.~N. Charette.
\newblock This car runs on code.
\newblock https://spectrum.ieee.org/this-car-runs-on-code - Accessed
  01-03-2022.

\bibitem{sandu2018new}
IA~Sandu, A~Salceanu, and O~Bejenaru.
\newblock New approach of the customer defects per lines of code metric in
  automotive sw development applications.
\newblock In {\em Journal of Physics: Conference Series}, volume 1065, page
  052006. IOP Publishing, 2018.

\bibitem{upstreamReport}
Upstream security's 2020 global automotive cybersecurity report.
\newblock
  https://upstream.auto/upstream-security-global-automotive-cybersecurity-report-2020/
  - Accessed 01-03-2022.

\bibitem{may20Incident}
Memory corruption vulnerability threatens smart vehicles.
\newblock https://upstream.auto/research/automotive-cybersecurity/?id=6150 -
  Accessed 01-03-2022.

\bibitem{10.5555/2671293.2671300}
Branden Ghena, William Beyer, Allen Hillaker, Jonathan Pevarnek, and J.~Alex
  Halderman.
\newblock Green lights forever: Analyzing the security of traffic
  infrastructure.
\newblock In {\em Proceedings of the 8th USENIX Conference on Offensive
  Technologies}, WOOT'14, page~7, USA, 2014. USENIX Association.

\bibitem{hahn2019security}
Dalton Hahn, Arslan Munir, and Vahid Behzadan.
\newblock Security and privacy issues in intelligent transportation systems:
  Classification and challenges.
\newblock {\em IEEE Intelligent Transportation Systems Magazine},
  13(1):181--196, 2019.

\bibitem{wp29}
Unece world forum for harmonization of vehicle regulations (wp.29).
\newblock https://unece.org/wp29-introduction - Accessed 01-03-2022.

\bibitem{abbaiot}
Abba-iot.
\newblock https://github.com/abbaiot - Accessed 01-03-2022.

\bibitem{Tuxera}
Tuxera.
\newblock Autonomous cars will generate more than 300 tb of data per year.
\newblock https://www.tuxera.com/blog/autonomous-cars-300-tb-of-data-per-year/
  - Accessed 01-03-2022.

\bibitem{GDPR1}
European Parliament and Council of~the European~Union.
\newblock On the protection of natural persons with regard to the processing of
  personal data and on the free movement of such data, and repealing directive
  95/46/ec (general data protection regulation).
\newblock Technical report, European Union, 2016.

\bibitem{Elephant}
P.~Hense.
\newblock They might be seen as green, but teslas are like elephants – they
  never forget.
\newblock
  https://www.jdsupra.com/legalnews/they-might-be-seen-as-green-but-teslas-81638/
  - Accessed 01-03-2022.

\bibitem{Chen}
S.~Chen, J.~Hu, Y.~Shi, Y.~Peng, J.~Fang, R.~Zhao, and L.~Zhao.
\newblock Vehicle-to-everything (v2x) services supported by lte-based systems
  and 5g.
\newblock {\em IEEE Communications Standards Magazine}, 2017.

\bibitem{Huang}
C.~Huang, H.~Wang, D.~Guo, G.~Zhang, G.~Xu, W.~Zhou, Y.~Cheng, Y.~Peng, K.~Xia,
  and F.~Lin.
\newblock A dynamic priority strategy for iov data scheduling towards key data.
\newblock {\em The Journal of Supercomputing}, 77, 2021.

\bibitem{Fraiji}
Y.~Fraiji, L.B. Azzouz, W.~Trojet, and L.A. Saidane.
\newblock Cyber security issues of internet of electric vehicles.
\newblock {\em 2018 IEEE Wireless Communications and Networking Conference
  (WCNC)}, 2018.

\bibitem{Lin}
J-R. Lin, T.~Talty, and O.K. Tonguz.
\newblock On the potential of bluetooth low energy technology for vehicular
  applications.
\newblock {\em IEEE}, 53(1):267--275, 2015.

\bibitem{ECHR}
European convention on human rights.
\newblock
  https://ec.europa.eu/digital-single-market/sites/digital-agenda/files/Convention\_ENG.pdf
  - Accessed 01-03-2022.

\bibitem{GDPREU}
GDPR.EU.
\newblock What is gdpr, the eu's new data protection law?
\newblock https://gdpr.eu/what-is-gdpr/ - Accessed 01-03-2022.

\bibitem{NYT}
Cambridge analytica and facebook: The scandal and the fallout so far.
\newblock
  https://www.nytimes.com/2018/04/04/us/politics/cambridge-analytica-scandal-fallout.html
  - Accessed 01-03-2022.

\bibitem{hgdpr}
European Data~Protection Supervisor.
\newblock The history of general data protection regulation.
\newblock
  https://edps.europa.eu/data-protection/data-protection/legislation/history-general-data-protection-regulation\_en
  - Accessed 01-03-2022.

\bibitem{CNIL}
The cnil’s restricted committee imposes a financial penalty of 50 million
  euros against google llc.
\newblock
  https://www.cnil.fr/en/cnils-restricted-committee-imposes-financial-penalty-50-million-euros-against-google-llc
  - Accessed 01-03-2022.

\bibitem{ICOBA}
British airways: Penalty notice.
\newblock https://ico.org.uk/action-weve-taken/enforcement/british-airways/ -
  Accessed 01-03-2022.

\bibitem{Dictionary}
Leeds University.
\newblock Intelligent transport systems.
\newblock
  http://www.its.leeds.ac.uk/projects/konsult/private/level2/instruments/instrument024/l2\_024a.htm
  - Accessed 01-03-2022.

\bibitem{Costantini2020}
Federico Costantini, Nikolas Thomopoulos, Fabro Steibel, Angela Curl, Giuseppe
  Lugano, and Tatiana Kováčiková.
\newblock Chapter eight - autonomous vehicles in a gdpr era: An international
  comparison.
\newblock {\em Advances in Transport Policy and Planning}, 5:191--213, 2020.

\bibitem{Delegated}
European Commission.
\newblock Commission delegated regulation (eu) 2015/962 of 18 december 2014
  supplementing directive 2010/40/eu of the european parliament and of the
  council with regard to the provision of eu-wide real-time traffic information
  services.
\newblock http://data.europa.eu/eli/reg\_del/2015/962/oj - Accessed 01-03-2022.

\bibitem{Livinglabs}
F.~Engels, A.~Wentland, and S.M. Pfotenhauer.
\newblock Testing future societies? developing a framework for test beds and
  living labs as instruments of innovation governance.
\newblock {\em Research Policy}, 48(9):103826, 2019.

\bibitem{AboutTesla}
Tesla.
\newblock Tesla’s mission is to accelerate the world’s transition to
  sustainable energy.
\newblock https://www.tesla.com/en\_GB/about - Accessed 01-03-2022.

\bibitem{ZapMap}
C.~Lilly.
\newblock Tesla model 3 dominates latest 2020 ev sales figures.
\newblock
  https://www.zap-map.com/tesla-model-3-dominates-latest-2020-ev-sales-figures/
  - Accessed 01-03-2022.

\bibitem{TeslaSupport}
Tesla.
\newblock Support: Car safety and security features.
\newblock
  https://www.tesla.com/en\_GB/support/car-safety-security-features\#:~:text=Sentry\%20Mode\%20is\%20a\%20feature,the\%20severity\%20of\%20the\%20threat
  - Accessed 01-03-2022.

\bibitem{CIP}
Steven Bliim.
\newblock What is sensitive data? sensitive data definition \& types.
\newblock https://cipherpoint.com/blog/what-is-sensitive-data/ - Accessed
  01-03-2022.

\bibitem{Jouvray}
C.~Jouvray, G.~Pellischek, and M.~Tigueracha.
\newblock Impact of a smart grid to the electric vehicle ecosystem from a
  privacy and security perspective.
\newblock {\em World Electric Vehicle Symposium and Exhibition (EVS27)}, pages
  1--10, 2013.

\bibitem{DEL}
University of~Delaware.
\newblock {\em Managing data confidentiality}.
\newblock https://www1.udel.edu/security/data/confidentiality.html - Accessed
  01-03-2022.

\bibitem{Pattinson}
J.A. Pattinson, H.~Chen, and S.~Basu.
\newblock Legal issues in automated vehicles: critically considering the
  potential role of consent and interactive digital interfaces.
\newblock {\em Human Soc Sci Commun}, 7(153), 2020.

\end{thebibliography}

\end{document}